\documentclass[lettersize,journal]{IEEEtran}
\usepackage{amsmath,amsfonts}
\usepackage{array}
\usepackage{textcomp}
\usepackage{siunitx}
\usepackage{enumitem}
\usepackage{stfloats}
\usepackage{subfig}
\usepackage{url}
\usepackage{verbatim}
\usepackage{graphicx}
\usepackage{cite}
\usepackage{bm}
\usepackage{stmaryrd}
\usepackage{booktabs}
\usepackage{multirow} 
\usepackage{amsthm}
\newtheorem{theorem}{Theorem}
\newtheorem{lemma}{Lemma}
\newtheorem{deft}{Definition}

\newtheorem{corollary}{Corollary}
\usepackage{amssymb}
\usepackage{amsmath}
\usepackage{amsmath,amssymb,amsfonts}
  
\usepackage{siunitx}
\usepackage{xcolor}
\definecolor{deepred}{RGB}{122,1,1}
\usepackage{makecell}
\usepackage{array} 
\usepackage{graphicx}
\newtheoremstyle{ieeeremark}
  {3pt}                    
  {3pt}                    
  {\normalfont}             
  {}                       
  {\bfseries}               
  {.}                       
  { }                       
  {}  
  
\theoremstyle{ieeeremark}
\newtheorem*{Rem}{Remark}

\usepackage{amsmath, amssymb, amsthm}

\theoremstyle{plain}
\usepackage[utf8]{inputenc} 
\usepackage[T1]{fontenc}
\usepackage{url}
\usepackage{ifthen}
\usepackage{amsfonts,algorithm,algpseudocode,xcolor}

\newcommand{\M}{\mathcal{M}}                 % modulo operator
\newcommand{\Dop}{\Delta}                    % difference operator
\newcommand{\Sop}{\mathrm{S}}                % shift operator
\newcommand{\norminf}[1]{\left\lVert #1 \right\rVert_{\infty}}
\newcommand{\sinc}{\operatorname{sinc}}

\newcommand{\OF}{\mathrm{OF}}

\begin{document}

\title{Difference-Based Recovery for Modulo Sampling: Tightened Bounds and Robustness Guarantees}

\author{Wenyi Yan, Zeyuan Li, Lu Gan, Honqing Liu, Guoquan Li
\thanks{
This work has been submitted to the IEEE for possible publication. Copyright may be transferred without notice, after which this version may no longer be accessible. 

Wenyi Yan and Lu Gan are with the College of Engineering, Design and Physical Sciences, Brunel University of London, U.K. Zeyuan Li is with the School of Integrated Circuits, Anhui University, China. Hongqing Liu and Guoquan Li are with the School of Communications and Information Engineering, Chongqing University of Posts and Telecommunications, China. }
\thanks{}}

% The paper headers
\markboth{DRAFT}%
{Shell \MakeLowercase{\textit{et al.}}: A Sample Article Using IEEEtran.cls for IEEE Journals}

\maketitle

\begin{abstract}

Conventional analog-to-digital converters (ADCs) clip when signals exceed their input range. Modulo (unlimited) sampling overcomes this limitation by folding the signal before digitization, but existing recovery methods are either computationally intensive or constrained by loose oversampling bounds that demand high sampling rates. In addition, none account for sampling jitter, which is unavoidable in practice. 
This paper revisits difference-based recovery and establishes new theoretical and practical guarantees. In the noiseless setting, we prove that arbitrarily high difference order reduces the sufficient oversampling factor from $2\pi e$ to $\pi$, substantially tightening classical bounds. For fixed order $N$, we derive a noise-aware sampling condition that guarantees stable recovery. For second-order difference-based recovery ($N=2$), we further extend the analysis to non-uniform sampling, proving robustness under bounded jitter. An FPGA-based hardware prototype demonstrates reliable reconstruction with amplitude expansion up to $\rho = 108$, confirming the feasibility of high-performance unlimited sensing with a simple and robust recovery pipeline.

\end{abstract}

\begin{IEEEkeywords}
Analog-to-digital converters (ADCs), modulo sampling, modulo ADCs, unlimited sampling, High-order differences, bandlimited signal reconstruction, field-programmable gate arrays (FPGAs).
\end{IEEEkeywords}

\section{Introduction}

Conventional analog-to-digital converters (ADCs) suffer from clipping when high-dynamic-range (HDR) signals exceed the ADC input range, resulting in information loss~\cite{koma_wide_2015, jo_very_2016,guarnieri_high_2011}. Modulo (``unlimited'') sampling systems~\cite{AB_Sampta,bhandari_unlimited_2018,bhandari_unlimited_2021} overcome this limitation by folding the input into a bounded interval $[-\lambda,\lambda)$ before digitization, then reconstructing the original signal. For a real-valued signal $g(t)$ with modulo threshold $\lambda > 0$, this modulo operation is defined as~\cite{bhandari_unlimited_2021}:
\begin{equation}
y(t)=\mathcal{M}_{\lambda}\bigl\{g(t)\bigr\}
    =g(t)-2\lambda\!\left\lfloor\frac{g(t)+\lambda}{2\lambda}\right\rfloor,
\label{eq:mod}
\end{equation}
where $\lfloor \cdot \rfloor$ denotes the floor function. 

Suppose that $g(t)$ has bandwidth $B$~\si{\hertz}, bounded amplitude $\|g\|_\infty$, and sampling frequency $f_s$~\si{\hertz}. We define the amplitude scaling factor $\rho$ and the oversampling factor $\mathrm{OF}$ as
\begin{equation}\label{eq:defrhoOF}
\rho \triangleq \frac{\|g\|_\infty}{\lambda}, \quad 
\mathrm{OF} \triangleq \frac{f_s}{2B}.
\end{equation}
Faithful recovery is achievable for $\rho \gg 1$, extending signal acquisition well beyond the native dynamic range of the ADC. Recent theoretical advances have established signal identifiability conditions and recovery algorithms for bandlimited signals, motivating hardware prototypes~\cite{bhandari_unlimited_2018,bhandari_unlimited_2021}.

Despite this progress, existing recovery methods either involve computationally intensive linear algebra operations (e.g., iterative least-squares or pseudoinverse optimisation)~\cite{Bhandari2022TSP_FP,Guo2023_ICASSP_ITERSIS,Shah2024_LASSO_B2R2} or require high oversampling factors~\cite{bhandari_unlimited_2021}. The classical high-order difference (HoD) method, although algorithmically simple, has a sufficient theoretical oversampling bound of $\mathrm{OF} \geq 2\pi e \approx 17.1$~\cite{bhandari_unlimited_2021}, and its noise sensitivity increases rapidly with the difference order $N$.

Another practical concern is that nearly all existing analyses assume perfectly uniform sampling. In real hardware, however, clock imperfections and analog front-end delays introduce \emph{timing jitter}, leading to non-uniform sampling instants. Such perturbations accumulate and can compromise theoretical guarantees if not explicitly accounted for. To the best of our knowledge, no prior work on unlimited sensing provides recovery guarantees under sampling jitter, leaving a critical gap between theory and practice.
This paper addresses both limitations through a joint algorithm–hardware design. The main contributions of this work are summarized as follows: 
\begin{itemize}
\item \textbf{Theoretical improvement}: 
We improve the sufficient oversampling condition for high-order differences (HoD), 
reducing it from $2\pi e$ to $\pi$ in the noiseless case, and derive a noise-aware 
sampling-rate condition for fixed order $N$.

\item \textbf{Second-order recovery for practical implementation}: 
As a special case of the fixed-$N$ analysis, we highlight the revised 
second-order difference (RSoD) scheme. This setting achieves a favorable trade-off 
between oversampling rate and noise robustness, making it attractive for practical use. 
In addition, we extend the $N=2$ analysis to non-uniform sampling and establish 
gap-dependent bounds that guarantee robustness under bounded timing jitter.

\item \textbf{Experimental validation}: 
We present a comprehensive evaluation of the RSoD algorithm through hardware 
measurements on the FPGA prototype, confirming robust reconstruction across the \si{\kilo\hertz} range with amplitude expansion up to $\rho=108$.
\end{itemize}

The remainder of this paper is organized as follows. Section~\ref{sec:liter} reviews the background and related work. Section~\ref{sec:theo} presents the tightened HoD analysis and a noise-aware sampling rule for fixed order. Section~\ref{sec:RSoD} develops the RSoD method and extends the analysis to non-uniform sampling with jitter. Section~\ref{sec;sim} presents simulation results and Section~\ref{sec:exp} evaluates the proposed reconstruction algorithm through hardware measurements. Section~\ref{sec:con} concludes the paper and outlines potential directions for future work.

\textbf{Notations:} Standard notation is adopted throughout: $\mathbb{N}$, $\mathbb{Z}$ and $\mathbb{R}$ denote natural numbers, integers and reals, respectively. The floor and ceiling functions are denoted $\lfloor\cdot\rfloor$ and $\lceil\cdot\rceil$ respectively. Continuous functions are denoted $x(t)$, $t\in\mathbb{R}$ with discrete counterparts $x[k]$, $k\in\mathbb{Z}$. The infinity norm is $\|x\|_\infty = \inf\{c_0 \geq 0 : |x(t)| \leq c_0\}$ for functions and $\|x\|_\infty = \max_k |x[k]|$ for sequences. 
The $N$-th order derivative is written as $x^{(N)}(t)$. $C^N(\mathbb{R})$ denotes $N$-times differentiable real-valued functions. For any discrete sequence $x[k]$, the first-order difference is given by 
\begin{equation}\label{eq:deffirstorder}
\left( {\Delta x} \right)\left[ k \right]= x\left[ {k + 1} \right] - x\left[ k \right]\end{equation}
and the $N^{\textrm{th}}$ order difference can be obtained by recursive application of the finite-difference operator 
\begin{equation}
{\Delta ^N}x= {\Delta ^{N - 1}}\left( {\Delta x}\right).
\end{equation}
The anti-difference operator is defined as  
\begin{equation}
\label{eq:AD}
\mathrm{S}(\left\{ {x\left[ k \right]} \right\}) = \sum\limits_{m = 1}^k x\left[ m \right]\quad k \in {\mathbb{Z}^ + }.
\end{equation}
A function $x(t) \in C^N(\mathbb{R})$ is $\Omega$-bandlimited ($\Omega=2\pi B$) if its Fourier transform satisfies $X(\omega) = \mathbf{1}_{[-\Omega,\Omega]}(\omega) X(\omega)$. $\mathrm{PW}_\Omega$ denotes the Paley-Wiener space of finite-energy, $\Omega$-bandlimited signals. Other system and performance-related notations are summarized in Table~\ref{table:notation}.

\begin{table}[!t]
\renewcommand{\arraystretch}{1.3}
\caption{Frequently Used Notations}
\label{table:notation}
\centering
\begin{tabular}{l|p{6.6cm}}
\toprule
\hline
\textbf{Notation} & \textbf{Explanation} \\
\hline
$\lambda$ & Modulo ADC threshold \\
\hline
$\mathcal{M}_{\lambda}\bigl\{\cdot\bigr\}$ & Modulo operation efined in Eq.~\eqref{eq:mod} \\
\hline
$x(t)$ & Noisy real Bandlimited input signal \\
\hline
$g(t)$ & Bandlimited continuous-time input signal \\
\hline
$\Omega$ & Signal bandwidth (\si{\radian\per\second})\\
\hline
$B$ & Signal bandwidth in \si{\hertz} ($\Omega/2\pi$) \\
\hline
$f_s$ & Sampling frequency\\\hline
$T$ & Sampling period $T=1/f_s$\\\hline
$t_k$ & $k$-th sampling instant; in the uniform case, $t_k=kT$\\ 
\hline
OF & Oversampling factor relative to Nyquist $\mathrm{OF}=f_s/(2B)$\\
\hline
$\rho=\|g\|_\infty/\lambda$ & Amplitude scaling factor \\
\hline
$y(t)$, $y[k]$ & Modulo signal $y(t)=\mathcal{M}_\lambda(g(t))$ and Modulo samples  \\
\hline
$\eta[k]$ & Additive noise in modulo samples\\
\hline
$y_\eta(t)$, $y_\eta[k]$ & continuous-time and discrete-time noisy modulo signals \\
\hline
$\varepsilon_{g}(t)$, $\varepsilon_{\gamma}[k]$ & Residual signal $\varepsilon_{g}(t)=g(t)-y(t)$ and residual samples\\
\hline
$\gamma[k]$, $\tilde{\gamma}[k]$ & Original samples $\gamma[k]=g(t_k)$ and reconstructed $\tilde{\gamma}[k]$ \\
\hline
$\nu, \mu_k$ & Sampling jitter parameters (jitter level and time offset) \\ \hline
$\sigma$ & Standard deviation of Gaussian noise \\ \hline
$b$ & Quantization resolution (bits/sample) \\ \hline
$\rho_\eta$ & Normalized noise level, $\rho_\eta = \|\eta\|_\infty / \lambda$ \\ \hline
$\mathrm{SNR}_r$ & Reconstruction signal-to-noise ratio \\ \hline
$\mathrm{PSNR}$ & Peak signal-to-noise ratio \\ \hline
$\mathrm{SINAD}$ & Signal-to-noise and distortion ratio \\ \hline
$\mathrm{ENOB}$ & Effective number of bits \\ \hline
%$\kappa_n$ & Integer alignment parameter in anti-differencing \\ \hline
\bottomrule
\end{tabular}
\end{table}

\begin{figure}[t]
    \centering
    \includegraphics[width=1\linewidth]{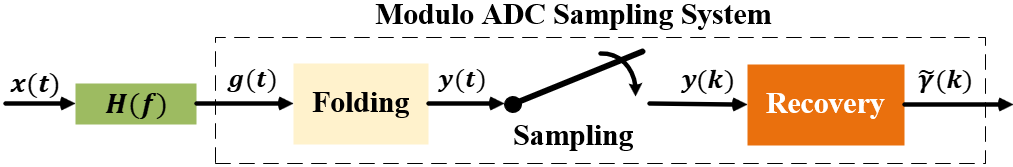}
\caption{Modulo ADC sampling system, where the input \(x(t)\) is prefiltered by \(H(f)\) to yield \(g(t)\), folded to produce \(y(t)\), sampled to obtain \(y(k)\), and digitally recovered as \(\tilde{\gamma}(k)\).}
    \label{fig:archi_pipeline}
\end{figure}

\section{Literature Review}
\label{sec:liter}
Consider a real bandlimited signal $x(t)$ observed in the presence of additive noise. 
As in conventional ADCs, an anti-aliasing filter $H(f)$ produces $g(t)$, 
which is subsequently processed by a modulo operation prior to sampling 
(Fig.~\ref{fig:archi_pipeline}). 
The bounded modulo output $y(t)$ is then sampled with period $T=1/f_s$, yielding 
$y[k]=\mathcal{M}_{\lambda}\!\left(g(kT)\right)$. 
A digital recovery algorithm reconstructs an estimate of $g(kT)$, denoted by $\tilde{\gamma}[k]$. 
Because the ADC processes only bounded signals, input peaks that would otherwise cause clipping 
are \emph{folded} and subsequently \emph{unwrapped} during recovery. 

The theoretical foundations of modulo sampling were established by Bhandari \textit{et al.}~\cite{bhandari_unlimited_2021}, who proved that finite-energy signals can be uniquely identified at any sampling rate strictly above Nyquist in the noiseless setting. Their high-order difference (HoD) method requires a sufficient oversampling factor of $\mathrm{OF}=2\pi e\approx 17.1$, with noise amplification that increases rapidly with the difference order $N$. Florescu \textit{et al.}~\cite{Florescu2022TSP_Hysteresis} showed that hardware hysteresis can yield robust recovery guarantees with relaxed sampling requirements. Romanov \textit{et al.}~\cite{Romanov2019_SPL_AboveNyquist} proposed a linear filtering approach that requires initial unfolded samples and is sensitive to noise, limiting practical applicability.

To improve robustness and efficiency, several algorithms exploit the piecewise-constant structure of the residue signal $\varepsilon_{g}(t)=g(t)-\mathcal{M}_{\lambda}(g(t))$. For discrete samples $\varepsilon_{\gamma}[k]=\varepsilon_{g}(kT)\in 2\lambda\mathbb{Z}$, the first-order difference $\Delta \varepsilon_{\gamma}[k]$ forms a sparse impulse train at folding instants~\cite{Bhandari2022TSP_FP,Azar2022_ICASSP_B2R2,Shah2024_LASSO_B2R2}. Algorithms such as Fourier–Prony~\cite{Bhandari2022TSP_FP}, residual-recovery (B$^2$R$^2$)~\cite{Azar2022_ICASSP_B2R2}, and LASSO-based variants~\cite{Shah2024_LASSO_B2R2} exploit this sparsity. However, as the amplitude expansion $\rho$ increases, fold density grows, weakening sparsity assumptions. FFT-based methods also suffer from spectral leakage on finite windows and ill-conditioning in pseudo-inverse steps, while time-domain approaches such as ITER-SIS~\cite{Guo2023_ICASSP_ITERSIS} mitigate these issues at increased computational cost.

Recent work has introduced sliding DFT methods with 1-bit side information~\cite{neil_side_24}, which can achieve lower mean square error (MSE) than conventional ADCs when $\mathrm{OF}>3$ and quantizer resolution $b>3$ bits/sample. However, theoretical conditions of $\mathrm{OF}>\rho+2$ still demand high oversampling for large $\rho$ (see Eq.~(9) in~\cite{neil_side_24}). Bandwidth-expansion strategies~\cite{zhu2025ironing} have also been explored, but typically require excessive oversampling and rely on models of analog folding circuits without fully accounting for ADC saturation effects.

Extensions of the modulo sampling framework include recovery of mixed sinusoids~\cite{Bhandari2022TSP_FP,zhang2024line,wang2025line,Guo_Spectral_2025}, finite rate-of-innovation signals, multi-channel architectures based on the Chinese remainder theorem (CRT)~\cite{Gan_Liu_MultiADC}, and 1-bit modulo ADCs~\cite{Graf_onebit_2019,onebit_2024,OneBits_2024}. While these represent important theoretical advances, they fall outside the present scope, which focuses on fundamental single-channel, memoryless modulo sampling for bandlimited signals of finite energy.

Despite these contributions, efficient recovery algorithms with provable performance bounds and low oversampling factors suitable for practice remain limited. 
Moreover, all existing works assume \emph{uniform sampling}, whereas in real-world hardware timing jitter is inevitable. 
To the best of our knowledge, no prior results establish recovery guarantees under non-uniform sampling. 
This motivates our analysis of second-order difference recovery with explicit gap-dependent bounds, showing that unlimited sensing remains stable under bounded sampling jitter.

\section{Difference-Based Recovery: Tighter Bounds and Fixed-Order Design}
\label{sec:theo}

\subsection{Problem statement}
Motivated by the need for low-complexity, real-time signal reconstruction, we revisit the HoD method~\cite{bhandari_unlimited_2021}. Consider a real-valued $g(t) \in \mathrm{PW}_{\Omega}$ with peak magnitude $\|g\|_{\infty} \leq \beta_g$ and sampling period $T=1/f_s$. Define the uniform samples and their modulo measurements as
 \begin{equation}\label{eq:defsamples}
\gamma[k]=g(kT), \quad y[k] = \mathcal{M}_\lambda(g(kT)).
\end{equation}
By definition of $\mathcal{M}_\lambda$ in~\eqref{eq:mod}, $\gamma[k]$ can be represented as
\begin{equation}
\gamma[k] = y[k] + \varepsilon_{\gamma}[k], \quad \text{where } \varepsilon_{\gamma}[k] \in 2\lambda \mathbb{Z}.
\end{equation}
The HoD reconstruction algorithm operates by computing $\Delta^{N}y$ and exploiting the key property that under appropriate sampling conditions,
\begin{equation}\label{eq:modcondition}
\Delta^{N}\gamma[k]=\mathcal{M}_{\lambda}(\Delta^{N}\gamma[k])=\mathcal{M}_{\lambda}(\Delta^{N}y[k]),
\end{equation}
which implies that 
\begin{equation}\label{eq:lift_levelN}
\mathcal{M}_{\lambda}(\Delta^{N}y[k]) - \Delta^{N}y[k] = \Delta^{N}\varepsilon_{\gamma}[k].
\end{equation}
The recovery is obtained by first calculating  $\Delta^{N}\varepsilon_{\gamma}[k]$ using ~\eqref{eq:lift_levelN}, followed by reconstruction of $\varepsilon_{\gamma}[k]$ via $N$ successive anti-difference operations with appropriate $2\lambda\mathbb{Z}$ alignment~\cite[Sec.~IV]{bhandari_unlimited_2021}. The original analysis provides sufficient conditions for noiseless recovery~\cite{bhandari_unlimited_2021}
\begin{align}
T &\leq \frac{1}{2\Omega e} \quad\Longleftrightarrow\quad \mathrm{OF} = \frac{f_s}{2B} > 2\pi e, \label{eq:BKRT_rate}\\
N &\geq \left\lceil \frac{\log \lambda - \log \|g\|_\infty}{\log(\Omega T e)} \right\rceil. \label{eq:BKRT_order}
\end{align}
In terms of amplitude scaling factor $\rho$ and oversampling rate $\mathrm{OF}$, the minimum order is
\begin{equation}\label{eq:Nmin-AB}
N_{\min} = \left\lceil \frac{\log(\|g\|_\infty/\lambda)}{\log(1/(\Omega Te))} \right\rceil = \left\lceil \frac{\log \rho}{\log\left(\mathrm{OF}/(\pi e)\right)} \right\rceil. 
\end{equation}
These conditions serve as our baseline. In the following analysis, we tighten these bounds by removing the factor $2e$ and derive practical fixed-order conditions that account for measurement noise.

\subsection{Revised Formula for Difference Order $N$}\label{sec:RDO}
In this section, we aim to cancel the constant $e$ in~\eqref{eq:BKRT_order}. This can be achieved by exploiting the mean-value theorem for divided differences~\cite{deBoor2005}, as shown in Lemma~\ref{lem:FD-bound}.  

\begin{lemma}[Forward\,-\,Difference Bound]\label{lem:FD-bound}
 Let $g(t)\in \mathrm{PW}_{\Omega}$ be $N$ times continuously differentiable.  Assume that $\gamma[k]$ is defined as in~\eqref{eq:defsamples} with a sampling period of $T$. 
For every integer $N\ge 1$ and $T>0$, the $N$‑th forward difference satisfies
\begin{equation}\label{eq:FD-bound}
  \bigl\lVert\Delta^{\,N}\gamma\bigr\rVert_{\infty}
   \;\le\; (T\,\Omega)^{N}\,\bigl\|g(t)\bigr\|_{\infty},
\end{equation}
\end{lemma}
\begin{IEEEproof}
The proof relies on the mean value theorem for divided differences, with the relevant preliminaries summarized in Appendix~\ref{sec:Apdif}. 
For a fixed $k\in\mathbb{Z}$, consider the $N$-th forward difference $\Delta^{N}\gamma[k]$. 
By the mean value theorem (see Prop.~22 in~\cite{deBoor2005} and Cor.~\ref{cor:forward-MVT} of the Appendix), there exists $\xi_k\in(kT,kT+NT)$ such that
\begin{equation}\label{eq:FD-MVT}
  \Delta^{N}\gamma[k] \;=\; T^{N} g^{(N)}(\xi_k).
\end{equation}
Since $g$ is bandlimited to $\Omega$ rad/s, the Bernstein inequality~\cite{bernstein1912ordre} yields
\begin{equation}\label{eq:Bernstein}
  \bigl|g^{(N)}(\xi_k)\bigr| \;\le\; \Omega^{N}\,\|g\|_{\infty}.
\end{equation}
Combining \eqref{eq:FD-MVT} and \eqref{eq:Bernstein} and taking the supremum over $k\in\mathbb{Z}$ establishes~\eqref{eq:FD-bound}.
\end{IEEEproof}

\begin{Rem}
In the original Lemma~2 of \cite{bhandari_unlimited_2021}, a looser upper bound was derived using a Stirling-type approximation for $N!$ to upper bound the $N$-th derivative norm. This introduced an additional multiplicative constant $e$, resulting in the bound $\|\Delta^{\,N} \gamma\|_\infty \le (e T \Omega)^N \bigl\|g(t)\bigr\|_\infty$. In contrast, our derivation applies the mean-value theorem for divided differences, which expresses the forward difference directly in terms of the exact $N$-th derivative at an intermediate point. This results in a tighter bound without the extra factor of $e$.
\end{Rem}

Following~\cite{bhandari_unlimited_2021}, condition~\eqref{eq:modcondition} is satisfied when $T < 1/\Omega$ and selecting $N$ sufficiently large such that
\[
\|\Delta^{N}\gamma\|_\infty \leq (\Omega T)^{N}\,\|g\|_\infty < \lambda.
\]
As $T\Omega  < 1$, the minimum integer $N$ can be expressed as 
\begin{equation}\label{eq:Nmin_new}
N_{\min} = \left\lceil \frac{\log(\|g\|_\infty/\lambda)}{\log(1/(\Omega T))} \right\rceil = \left\lceil \frac{\log \rho}{\log(\mathrm{OF}/\pi )} \right\rceil. 
\end{equation}
Compared to the baseline in~\eqref{eq:Nmin-AB}~\cite{bhandari_unlimited_2021}, our bound eliminates the constant $e$, strictly reducing $N_{\min}$ for a given $\mathrm{OF}$. This yields two key advantages: (\emph{i}) lower implementation cost via fewer (anti-)~differencing stages, and (\emph{ii}) improved robustness by curbing noise amplification (which scales as $2^N$~\cite[Theorem~3]{bhandari_unlimited_2021}). Section~\ref{sec;sim} validates these gains through simulations.

\subsection{Fixed-Order Difference Recovery Algorithm}\label{sec:RSoD_Rec}

\begin{algorithm}[t]
\caption{Revised Fixed-Order Difference Recovery from Noisy Modulo Samples}\label{alg:noisy_RSoD}
\begin{algorithmic}[1]
\Require $y_\eta[k]$, $\lambda$ and $\beta_g \ge \norminf{g}$.
\Ensure $\tilde g(t)\approx g(t)$.
\State Compute $\bigl(\Dop^{N} y_\eta\bigr)[k]$.
\State Compute $\bigl(\Dop^{N}\varepsilon_\gamma\bigr)[k]
      =\bigl(\M_{\lambda}(\Dop^{N}y_\eta)-\Dop^{N}y_\eta\bigr)[k]$.
 \State     Set $s_{(0)}[k]=\bigl(\Dop^{N}\varepsilon_\gamma\bigr)[k]$.
\For{$n=0 : N-2$} \label{line:for}
  \State \quad $s_{(n+1)}[k]=\bigl(\Sop s_{(n)}\bigr)[k]$.
  \State \quad $s_{(n+1)} =
      2\lambda \left\lfloor \dfrac{s_{(n+1)}/\lambda}{2} \right\rceil$
  \State \quad \textcolor{blue}{$\displaystyle J = \Bigl\lceil 4\Bigl( \dfrac{\beta_g}{\lambda}+2^{N-2} \Bigr) \Bigr\rceil$} 
  \Statex \vspace{0.1em}
  \State  \quad 
  \textcolor{blue}{$
\kappa_n =
\left\lfloor
\frac{ \bigl( \mathrm{S}^2 \Delta^{n} \varepsilon_{\gamma} \bigr)[1]
     - \bigl( \mathrm{S}^2 \Delta^{n} \varepsilon_{\gamma} \bigr)[J+1]}
     {2J\lambda}
+ \frac{1}{2}
\right\rfloor.
$}
\Statex \vspace{0.1em}
  \State \quad $s_{(n+1)}[k]=s_{(n+1)}[k]+2\lambda\,\kappa_n$.
\EndFor
\State $\tilde{\gamma}[k]=\bigl(\Sop s_{(N-1)}\bigr)[k]+y_\eta[k]+2m\lambda,\ \ m\in\mathbb{Z}$.
\State Compute $\tilde g(t)$ from $\tilde{\gamma}[k]$ via low-pass filter interpolation: 
       $\displaystyle \tilde g(t)=
       \sum_{k\in\mathbb{Z}}\tilde{\gamma}[k]\;\sinc\!\bigl(t/T-k\bigr)$.
\end{algorithmic}
\end{algorithm}

In this section, we modify the HoD algorithm to achieve reconstruction at lower sampling rates. Consider noisy modulo measurements
\begin{equation}\label{eq:noisy-measurements}
y_\eta[k] = \mathcal{M}_{\lambda}(\gamma[k])+\eta[k]=y[k]+\eta[k],
\end{equation}
where $\eta[k]$ represents additive noise with bounded magnitude $\|\eta\|_{\infty} \leq \epsilon$. As established in~\cite{bhandari_unlimited_2021}, recovering $\gamma[k]$ is equivalent to recovering the residue sequence $\varepsilon_{\gamma}[k]$. Since $\Delta^{n} \varepsilon_{\gamma} \in 2\lambda\mathbb{Z}$ for $1 \leq n \leq N$, this discrete constraint significantly reduces the ill-conditioning compared to direct recovery of $\gamma[k]$ from the noisy modulo samples $y_\eta[k]$.

For fixed difference order $N$, Algorithm~\ref{alg:noisy_RSoD} presents our revised fixed order difference approach, with key changes highlighted in blue color. The overall pipeline follows the classical HoD framework~\cite{bhandari_unlimited_2021}: compute $\Delta^N y_\eta$, unwrap at the difference level to obtain $\Delta^N\varepsilon_\gamma$, then perform anti-differencing with $2\lambda\mathbb{Z}$ alignment.

The main difference is an adaptive block length $J$ for the rounding operation. While the baseline method~\cite{bhandari_unlimited_2021} employs an order-independent window $J = \lceil 6\beta_g/\lambda \rceil$, we use an $N$-dependent choice:
\begin{equation}\label{eq:proposedJ}
J = \left\lceil 4\left(\frac{\beta_g}{\lambda} + 2^{N-2}\right) \right\rceil.
\end{equation}
This selection accounts for the growth of $\Delta^N$ operations and maintains the $\frac{1}{4}$ rounding guard at each anti-differencing level.

In the noiseless case, this refinement combined with Lemma~\ref{lem:FD-bound} eliminates the extra factor of 2, improving the sufficient condition from $T \leq 1/(2\Omega e)$ to $T < 1/\Omega$ (equivalently, $\mathrm{OF} > \pi\rho^{1/N}$). Furthermore, Theorem~\ref{thm:noisy} demonstrates that this algorithm extends to bounded noise scenarios with explicit oversampling requirements.

\begin{theorem}[Noisy Recovery for Revised Fixed-Order Difference Algorithm]\label{thm:noisy}
Let $g \in \mathrm{PW}_{\Omega}$ be a bandlimited signal with $\|g\|_{\infty} < \infty$, and let $\gamma[k] = g(kT)$ denote its discrete samples. Consider noisy modulo measurements $y_\eta[k]$ as defined in~\eqref{eq:noisy-measurements}.

For fixed difference order $N \geq 1$, if the condition
\begin{equation}\label{eq:noisy_reccond}
(\Omega T)^{N}\,\|g\|_\infty + 2^{N}\,\|\eta\|_\infty < \lambda
\end{equation}
is satisfied, then Algorithm~\ref{alg:noisy_RSoD} produces
\[
\tilde{\gamma}[k] = \gamma[k] + \eta[k] + 2m\lambda, \quad m \in \mathbb{Z}.
\]

Defining the normalized error $\rho_\eta = \|\eta\|_\infty/\lambda$, condition~\eqref{eq:noisy_reccond} is equivalent to the oversampling requirement
\begin{equation}\label{eq:OF_noisy}
\mathrm{OF} = \frac{f_s}{2B} > \pi\left(\frac{\rho}{1-2^{N}\rho_\eta}\right)^{\!1/N}, \quad (2^{N}\rho_\eta < 1).
\end{equation}

In the noiseless case ($\eta \equiv 0$, i.e., $\rho_\eta = 0$), this reduces to
\[
\mathrm{OF} > \pi\,\rho^{1/N}.
\]
As $N \to \infty$, the required $\mathrm{OF}$ approaches $\mathrm{OF} > \pi$.
\end{theorem}

The proof is provided in Appendix~\ref{Asec:noisy_bound}.

\textbf{Comparisons with~\cite{bhandari_unlimited_2021}:} 
In the noisy setting, \cite[Theorem~3]{bhandari_unlimited_2021} admits bounded perturbations provided that
\[
\rho_\eta \;<\; \tfrac{1}{4}\,(2\rho)^{-1/\alpha}, \qquad \alpha\in\mathbb{N},
\]
with a sampling condition
\begin{equation}\label{eq:baseOF_noisy}
  T \;\le\; \frac{1}{2^{\alpha}\,\Omega\,e}
\quad\Longleftrightarrow\quad
\mathrm{OF}\;\ge\;2^{\alpha}\,\pi e,  
\end{equation}
and a difference order choice
$N \;=\; \left\lceil \frac{\log\rho + 1}{\log\!\big(\mathrm{OF}/(\pi e)\big)} \right\rceil$.

Hence, to keep the sampling rate \emph{independent of} \(\rho\), their design lets \(N\) grow with \(\rho\). Meanwhile, \(\mathrm{OF}\) in~\eqref{eq:baseOF_noisy} grows exponentially with $\alpha$ \(\{2^{\alpha}\pi e:\alpha=1,2,\dots\}\) (e.g., \(2\pi e\!\approx\!17.1\), \(4\pi e\!\approx\!34.2\), \(8\pi e\!\approx\!68.4\)), offering limited granularity. By contrast, we \emph{fix} a small, practical order \(N\) and obtain a \emph{noise-aware}, continuous-rate condition of $\mathrm{OF}$ in~\eqref{eq:OF_noisy}. Thus, \(N\) does \emph{not} grow with \(\rho\); the sampling rate scales mildly as \(\rho^{1/N}\) and can be tuned finely to the measured noise level. 

Table~\ref{tab:OF_comparison} illustrates the oversampling requirements at 
moderate noise levels ($\rho_\eta=0.10{:}0.02{:}0.20$) for $\rho=10$. 
Our fixed-order bounds with $N=2,3$ are consistently one to two orders of magnitude 
smaller than those implied by Theorem~3 of~\cite{bhandari_unlimited_2021}. 
Even after removing the extraneous factor $e$ (cf. Lemma~\ref{lem:FD-bound}), the bound derived in~\cite{bhandari_unlimited_2021} remains 
discrete in $\alpha$ and grows exponentially, leading to much larger oversampling 
factors. By contrast, our analysis yields a continuous condition in $\rho_\eta$ that 
closely tracks practical operating points: while higher $N$ reduces the rate when 
$\rho_\eta$ is extremely small (e.g., $\rho_\eta<0.01$), at realistic noise levels, $N=2$ offers the 
most favourable trade-off between robustness and sampling rate.

\renewcommand{\arraystretch}{1.3}
\begin{table}[t]
\centering
\caption{Comparison of oversampling factors $\mathrm{OF}$ for $\rho=10$ at different noise levels $\rho_\eta$. 
Our bounds are shown for fixed $N=2,3$, while Theorem~3 of~\cite{bhandari_unlimited_2021} yields discrete thresholds 
($2^\alpha \pi e$) and the variant without $e$ follows from our Lemm~\ref{lem:FD-bound}.}
\label{tab:OF_comparison}
\begin{tabular}{c|cc|cc}
\toprule
\hline
\multirow{2}{*}{$\rho_\eta$} 
& \multicolumn{2}{c|}{Our Bounds} 
& \multicolumn{2}{c}{Bounds of~\cite{bhandari_unlimited_2021}} \\
\cline{2-5}
 & $N=2$ & $N=3$ 
 & Original & w/o $e$  \\
\hline
0.10 & 12.83 & 11.57 & 136.64 & 50.27 \\
\hline
0.12 & 13.78 & 19.70 & 273.27 &  100.53 \\
\hline
0.14 & 14.98 & - &   564.54 & 201.06\\  
\hline
0.16 & 16.56 & - & 1093.09  & 402.12 \\
\hline
0.18 & 18.77 & - & 8744.69  & 3216.99 \\
\hline
0.20 & 22.21 & - & 139915.01 &  51472.85\\
\hline
\bottomrule
\end{tabular}
\end{table}

%\vspace{0.4em}
\subsection{Analysis under Quantization Noise}
The following Corollary investigates Algorithm~\ref{alg:noisy_RSoD}'s performance under quantization noise. 
\begin{corollary}[Quantized Modulo ADC, $b$-bit]
\label{cor:quantized}
For a uniform $b$-bit ADC quantizing modulo samples over $[-\lambda,\lambda)$, the sufficient oversampling factor is
\begin{equation}\label{eq:quant-OF}
\mathrm{OF} > \pi\left(\frac{\rho}{1 - 2^{N-b}}\right)^{\!1/N}\quad (b>N).
\end{equation}
%valid when $b > N$.
\end{corollary}

\begin{IEEEproof}
With quantization step $q = 2\lambda/2^b$, the quantization error satisfies $\|\eta\|_\infty \leq q/2 = \lambda 2^{-b}$, yielding $\rho_\eta \leq 2^{-b}$. Substitution into \eqref{eq:OF_noisy} of Theorem~\ref{thm:noisy} produces \eqref{eq:quant-OF}, where $b > N$ ensures $2^N\rho_\eta < 1$.
\end{IEEEproof}

The following corollary compares modulo and conventional ADCs in terms of 
the \emph{signal-to-noise and distortion ratio} (SINAD) and the 
\emph{effective number of bits} (ENOB).

\begin{corollary}[Conventional ADC Comparison]
\label{cor:conventional}
Consider a conventional $b_c$-bit ADC covering $[-\|g\|_\infty,\|g\|_\infty]$ 
and a modulo ADC with $b_m$ bits over $[-\lambda,\lambda)$. 
At an equal nominal bit depth, the modulo ADC achieves
\begin{align}
\mathrm{SINAD}_{\mathrm{mod}} &\approx \mathrm{SINAD}_{\mathrm{conv}} + 20\log_{10}\rho, \\
\mathrm{ENOB}_{\mathrm{mod}} &\approx \mathrm{ENOB}_{\mathrm{conv}} + \log_2 \rho.
\end{align}
Equivalently, $b_m$-bit modulo ADCs match $b_c = b_m + \log_2\rho$.
\end{corollary}
\begin{IEEEproof}
Comparing $q_c=2\rho\lambda/2^{b_c}$ and $q_m=2\lambda/2^{b_m}$ gives a noise ratio $\rho^2$, yielding the SINAD/ENOB relations.
\end{IEEEproof}

\vspace{0.1em}

\textbf{Comparison with LASSO-based recovery~\cite{kvich_practical_2025}}:  
Lemma~1 in~\cite{kvich_practical_2025} requires $\mathrm{OF}$ $> \rho + 2$ (linear in $\rho$) for LASSO recovery with 1-bit side information (see Eq.~(9) in.~\cite{kvich_practical_2025}). In contrast, Corollary~\ref{cor:quantized} yields $\mathrm{OF}=O\!\left(\rho^{1/N}\right)$ for fixed $N$ and $b > N$, which is substantially weaker in $\rho$-dependence while maintaining a clear noise-limited stability region.

\section{Revised Second order Difference Method}
\label{sec:RSoD}
\subsection{Esimation of OF}
Theorem~\ref{thm:noisy} makes the rate–robustness trade-off explicit: in the noiseless limit, larger $N$ lowers the required rate as $\mathrm{OF}>\pi\,\rho^{1/N}$. With bounded perturbations, the feasibility region $2^{N}\rho_\eta<1$ shrinks exponentially in $N$. In the quantized case, one also needs $b>N$. 

Balancing these effects, we \emph{fix $N{=}2$} and refer to Algorithm~\ref{alg:noisy_RSoD} specialized to $N{=}2$ as the \textbf{RSoD} algorithm. RSoD is used throughout our hardware experiments to unwrap the modulo samples. In this case, \eqref{eq:OF_noisy} reduces to
\begin{equation}\label{eq:RSoD_OF}
\mathrm{OF}>\pi\!\left(\frac{\rho}{1-4\rho_\eta}\right)^{1/2},\qquad \rho_\eta<\tfrac14,
\end{equation}
which provides comfortable noise tolerance. In the quantized setting, this implies only $b\ge3$ bits/sample.

It should be pointed out that the generic bound $\|\Delta_T^{N}\gamma\|_\infty \le (T\Omega)^{N}\|g\|_\infty$ in Lemma~\ref{lem:FD-bound} can be sharpened for specific functions $g(t)$. For sinc functions with $\Omega = 2\pi B$, we have $\|g^{(N)}\|_\infty \le \frac{\Omega^N}{N+1}\|g\|_\infty$~\cite{gronwall1913,dunkel1920}. With $N=2$,~\eqref{eq:RSoD_OF} becomes
\begin{equation}\label{eq:OFRSOD_sinc}
\mathrm{OF} > \pi\left(\frac{\rho}{3(1-4\rho_\eta)}\right)^{1/2} \quad \rho_\eta < \tfrac{1}{4},
\end{equation}
providing a $\sqrt{3}$ reduction in $\mathrm{OF}$. For near-brickwall spectra,~\eqref{eq:OFRSOD_sinc} yields a tighter bound than~\eqref{eq:RSoD_OF}, as confirmed by experiments in Section~\ref{sec:exp}.

\subsection{Jittering Effect}
We now consider the impact of Jittering (or more generally, non-uniform sampling) for reconstruction from modulo samples. Classical nonuniform sampling theory guarantees exact recovery of all bandlimited 
signals when the sampling perturbation satisfies $|t_k-kT|<T/4$ for all $k$ 
(Kadec’s $1/4$ theorem~\cite{wave_non,non_ka}). It is interesting to consider how we can generalize this result to modulo samples. The following theorem presents the OF bound:  

\begin{theorem}[Noisy Recovery with Jitter for RSoD]
\label{thm:noisy_N2_jitter}
Let $g\in \mathrm{PW}_\Omega$ with $\|g\|_\infty<\infty$, and let the sampling times be
$t_k=kT+\mu_k$ with $|\mu_k|<\nu T$ for all $k$ (jitter level $\nu\ge 0$).
Define samples $\gamma[k]=g(t_k)$ and noisy modulo measurements
$y_\eta[k]=\mathcal{M}_\lambda(\gamma[k]+\eta[k])$, where $\|\eta\|_\infty<\infty$.
Set the normalized quantities $\rho:=\|g\|_\infty/\lambda$ and
$\rho_\eta:=\|\eta\|_\infty/\lambda$. If the following condition is satisfied 
\begin{equation}
\label{eq:cond_N2_jitter_updated}
\rho\big((T\Omega)^2+4\nu T\Omega\big) + 4\rho_\eta < 1\quad\big(\rho_\eta<\tfrac14\big),
\end{equation}
then the RSoD (Algorithm~\ref{alg:N2_jitter})
produces
\[
\tilde{\gamma}[k] \;=\; \gamma[k]+\eta[k]+2m\lambda, 
\qquad m\in\mathbb{Z},
\]
i.e., exact unwrapping up to a global $2\lambda$ ambiguity. Equivalently, the admissible range of $T\Omega $ is 
\begin{equation}
\label{eq:x_bound_N2_jitter_updated}
T\Omega< 
 -2\nu + \sqrt{4\nu^2+\tfrac{1-4\rho_\eta}{\rho}}.
\end{equation}
In terms of $\OF$, this is equivalent to
\begin{equation}\label{eq:OF_jitter}
\mathrm{OF} \;>\; \frac{\pi}{ -2\nu + \sqrt{4\nu^2+\frac{1-4\rho_\eta}{\rho}}}. 
\end{equation}
\end{theorem}
The proof is provided in Appendix~\ref{Asec:Jitter}. For a Sinc function with bandwidth of $\Omega$, the Bernstein-type bounds refine to
$\|g^{(2)}\|_\infty \le \frac{\Omega^2}{3}\|g\|_\infty$ and $\|g^{(1)}\|_\infty \le \frac{\Omega}{2}\|g\|_\infty$. Hence, the jitter-aware condition aware condition of~\eqref{eq:cond_N2_jitter_updated} becomes
\[
\frac{\rho}{3}\,(T\Omega)^2 + 2\rho\nu\,T\Omega + (4\rho_\eta - 1) < 0. 
\]
Solving the quadratic inequality for $T\Omega$ yields the admissible upper bound as the smaller root of the equation, implying
\[
T\Omega < -3\nu + \sqrt{3}\,\sqrt{\,3\nu^2 + \frac{1-4\rho_\eta}{\rho}\,},
\]
which produces
\begin{equation}\label{eq:jitter-sinc}
\OF\ge \frac{\pi}{-3\nu + \sqrt{3}\sqrt{3\nu^{2} + \frac{1 - 4\rho_{\eta}}{\rho}}}.
\end{equation}

\begin{Rem}[Effect of Jitter]
When $\nu=0$, Theorem~\ref{thm:noisy_N2_jitter} reduces exactly to the 
uniform-sampling result in Theorem~\ref{thm:noisy}. 
For small jitter levels ($\nu \ll 1$), a Taylor expansion of the admissible bound 
shows that the right hand side of~\eqref{eq:x_bound_N2_jitter_updated} can be approximated as $-2\nu +\sqrt{\tfrac{1-4\rho_\eta}{\rho}}$
i.e., timing jitter effectively reduces the permissible oversampling budget 
by a linear penalty proportional to $\nu$. 
Since the dominant term in the recovery condition remains $(T\Omega)^2$, 
the degradation due to jitter is minor in high-oversampling regimes. 
This explains why difference-based unlimited sensing is robust to modest 
timing jitter in practice.
\end{Rem}

\textbf{Potential Extensions}:
Our framework for modulo recovery with sampling jitter also applies to frequency-sparse signal models. We demonstrate that the key unwrapping step is valid even on non-uniform grids. This enables a new approach: unwrap modulo samples acquired at perturbed times, then use standard sparse recovery methods (e.g., atomic norm minimization) for spectral estimation. This points to significant potential for our technique in compressed sensing and randomized sampling.

\begin{algorithm}[t]
\caption{RSoD ($N{=}2$) with Jitter: Unwrapping + Non-Uniform Reconstruction}
\label{alg:N2_jitter}
\begin{algorithmic}[1]
\Require Noisy modulo samples $y_\eta[k]$, threshold $\lambda$, magnitude prior $\beta_g\!\ge\!\|g\|_\infty$, and \emph{time instants} $\{t_k\}$ with $t_k=t_0+kT+\mu_k$.
\Ensure $\tilde g(t)\approx g(t)$.
\State Compute $\Delta^{2} y_\eta[k]$.
\State Compute $\Delta^{2}\varepsilon_\gamma[k]
 \leftarrow \mathcal{M}_\lambda(\Delta^{2}y_\eta[k])-\Delta^{2}y_\eta[k]$.
%\State $s_{(0)} \leftarrow \Delta^{2}\varepsilon_\gamma$.
\State $s_{(1)} \leftarrow \mathrm{S}\Delta^{2}\varepsilon_\gamma$ %\Comment{one antidifference}.
\State Integer alignment: $s_{(1)} \leftarrow 2\lambda\Big\lfloor \tfrac{s_{(1)}}{2\lambda} \Big\rceil$.
\State Estimate $\kappa_1$ as in Alg.~\ref{alg:noisy_RSoD}; 
\State Set $s_{(1)} \leftarrow s_{(1)} + 2\lambda\,\kappa_1$.
\State Recover unfolded samples: $\tilde{\gamma}[k] \leftarrow \mathrm{S}\,s_{(1)} + y_\eta[k] + 2m\lambda,\; m\in\mathbb{Z}$.
\State \textbf{Non-uniform bandlimited recovery from }$\{(t_k,\tilde{\gamma}[k])\}$\textbf{:} use any standard method (e.g., NUFFT-based least squares, frame/Paley–Wiener methods, or kernel interpolation) to obtain $\tilde g(t)$.
\end{algorithmic}
\end{algorithm}

\section{Simulations}\label{sec;sim}
In this section, we compare the proposed RSoD method with the classical HoD algorithm in~\cite{bhandari_unlimited_2021} as well as with state-of-the-art approaches, and further validate the theoretical bounds derived in Section~\ref{sec:theo} and ~\ref{sec:RSoD}.

\begin{figure}[t]
    \centering
    \includegraphics[width=0.8\linewidth]{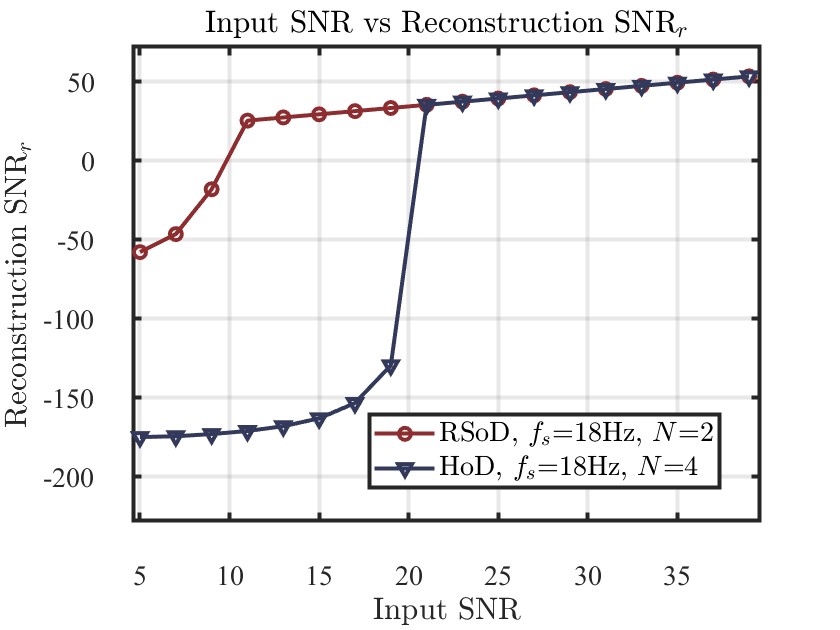}
\caption{Reconstruction $\mathrm{SNR}_r$ versus input SNR for a $\mathrm{sinc}(2Bt)$ input with $B=\SI{0.5}{\hertz}$ and $\rho=12$. 
Both methods operate at $f_s=\SI{18}{\hertz}$ (18$\times$ Nyquist), with RSoD using $N=2$ and HoD using $N=4$.}
  \label{fig:snr_vs_snr}
\end{figure}

\textbf{Simulation 1: Baseline Comparison (RSoD vs. HoD)}. This simulation benchmarks the proposed RSoD algorithm against the original HoD method~\cite{bhandari_unlimited_2021}. 
A sinc input $g(t)=\mathrm{sinc}(2Bt)$ with $B=\SI{0.5}{\hertz}$ ($f_{\mathrm{Nyq}}=\SI{1}{\hertz}$) and $\rho=12$ is sampled at $f_s=\SI{18}{\hertz}$ ($\mathrm{OF}=18$). 
Based on~\eqref{eq:Nmin-AB}, the HoD algorithm requires $N=4$ for recovery. 
Gaussian noise is added to the modulo samples, and performance is evaluated by the reconstruction SNR (SNR$_r$), defined as
\begin{equation}
\text{SNR}_r = 10 \log_{10} 
\frac{\sum_{k} |\gamma(k)|^2}{\sum_{k} |\gamma(k)-\tilde{\gamma}(k)|^2}.
\label{eq:rrse_def}
\end{equation}
Fig.~\ref{fig:snr_vs_snr} shows that RSoD achieves reliable reconstruction at an input SNR of about \SI{11}{\decibel}, compared to \SI{21}{\decibel} for HoD, indicating significantly improved noise resilience.

\begin{figure*}[t]
    \centering
    \subfloat[]{
        \includegraphics[width=0.8\linewidth]{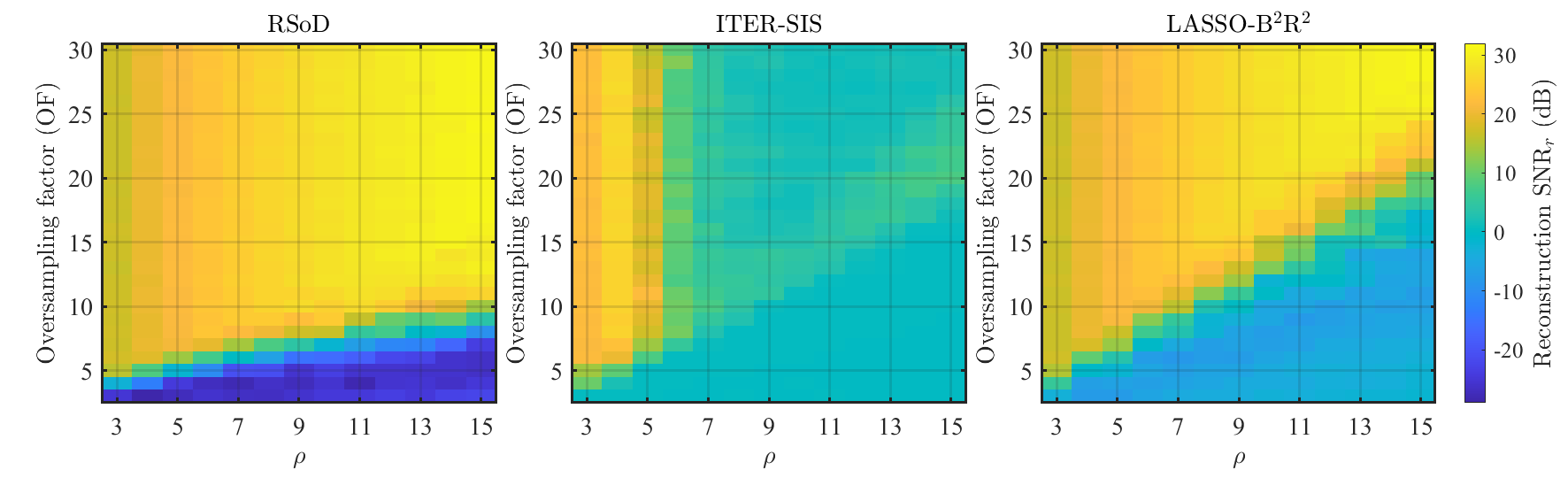}
        \vspace{-0.2cm}
    }\\
    \subfloat[]{
        \includegraphics[width=0.8\linewidth]{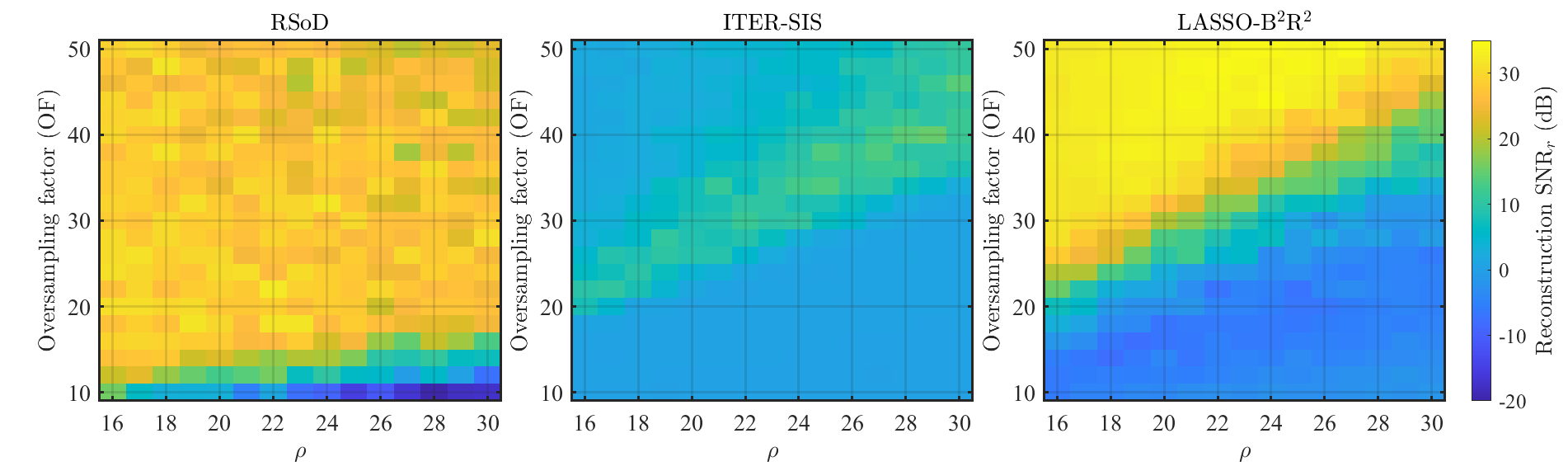}
        \vspace{-0.2cm}
    }
\caption{Reconstruction SNR under Gaussian noise for RSoD, ITER--SIS, and
LASSO--B$^2$R$^2$. \textbf{(a)} $\rho\in [3, 15]$ and $\mathrm{OF}\in[3,30]$ and input $\mathrm{SNR}=19.40$ dB and $b=3$. 
\textbf{(b)} $\rho\in [16:30]$ and $\mathrm{OF}\in[10:2:50]$ and input $\mathrm{SNR}=15$ dB and $b=4$.}
    \label{fig:std_fixed}
\end{figure*}

\textbf{Simulation 2: Comparison with Existing Recovery Algorithms.} 
We benchmark against state-of-the-art algorithms \textbf{ITER-SIS}~\cite{Guo2023_ICASSP_ITERSIS} and \textbf{LASSO-B$^2$R$^2$}~\cite{Shah2024_LASSO_B2R2} using their default settings (up to 100 iterations for ITER-SIS and 1000 for LASSO-B$^2$R$^2$). Alternative algorithms are excluded due to practical limitations: linear-prediction methods require unfolded initialisation samples~\cite{Romanov2019_SPL_AboveNyquist}; Fourier–Prony suffers from spectral leakage and fold-count dependence~\cite{Bhandari2022TSP_FP,Guo2023_ICASSP_ITERSIS,Shah2024_LASSO_B2R2}; one-bit side-information methods are incompatible with modulo-only setups~\cite{kvich_practical_2025}; and hysteresis-assisted schemes require nonlinearity calibration beyond our scope~\cite{Florescu2022TSP_Hysteresis}.

We use bandlimited signals with maximum frequency $B=0.5$~Hz ($f_{\mathrm{Nyq}}=1$~Hz). The observation interval is $25$~s, providing sufficiently many samples for reliable recovery analysis across different oversampling factors. Each test signal is generated as a linear combination of six shifted sinc functions, 
\begin{equation}
\label{eq:sincSim}
x(t) \;=\; \sum_{i=1}^{6} a_i \, \sinc(t - i),
\end{equation}
where the coefficients $a_i \sim \mathcal{U}[-1,1]$ (uniform distribution).
In our experiments, modulo samples are corrupted by Gaussian noise and quantized to $b$ bits per sample. Fig.~\ref{fig:std_fixed}(a) shows reconstruction SNR$_r$ phase transitions for $\rho\in[3,15]$ and $\mathrm{OF}\in[6,30]$ with input SNR of 19.40\,dB (from modulo hardware measurements in Fig.~6 of~\cite{Guo_6K_2024}) and $b=3$ bits (minimum required by RSoD theoretical analysis). Fig.~\ref{fig:std_fixed}(b) extends to $\rho\in[16,30]$ and $\mathrm{OF}\in[10,50]$ at 15\,dB input SNR with $b=4$ bits, where the lower SNR reflects increased folding difficulty at larger $\rho$. 
Each result is averaged over 100 independent trials per $(\rho,\text{OF})$ pair.

In the high-SNR regime (19.40\,dB, Fig.~\ref{fig:std_fixed}(a)), RSoD achieves consistently high SNR$_r$ across a wide range of $\rho$ with low OF ($\mathrm{OF}\approx 8$–10). The empirical transition boundary is consistent with theoretical scaling $\mathrm{OF} = \mathcal{O}(\sqrt{\rho})$, confirming that RSoD can reliably unwrap dense folds with moderate oversampling. ITER-SIS excels for $\rho\leq6$ but degrades at larger $\rho$ due to its annihilating filter requiring accurate fold-count $K$ estimation. As $\rho$ increases, $K$ grows and the Prony step becomes ill-conditioned. LASSO-B$^2$R$^2$ demonstrates better stability but requires oversampling scaling roughly linearly with $\rho$, consistent with compressed-sensing sparsity bounds.

In the lower-SNR regime (15\,dB, Fig.~\ref{fig:std_fixed}(b)), RSoD shows degradation in average SNR$_r$ when noise amplitude exceeds the normalized error threshold, as error propagation amplifies through anti-difference operations. LASSO-B$^2$R$^2$, while less efficient in sampling rate, appears more stable under noisy conditions due to its $\ell_1$-sparse optimisation. ITER-SIS continues degrading with larger $\rho$ from ill-conditioned Vandermonde systems. Overall, RSoD is most sample-efficient in moderate-to-high SNR settings. But at lower SNRs, LASSO-B$^2$R$^2$ offers more stable recovery at significantly higher oversampling cost (under the tested hyperparameters). ITER-SIS provides excellent performance for $\rho\leq 6$ but degrades substantially as $\rho$ increases due to numerical instabilities.

Table~\ref{tab:alg_comparison_time_sim} reports average execution times. RSoD is consistently fastest, requiring $<2$\,ms even at $(\rho,\mathrm{OF}){=}(15,30)$. ITER-SIS shows moderate scaling from ${\approx}14$\,ms at $(3,5)$ to $>50$\,ms at $(15,30)$. LASSO-B$^2$R$^2$ requires more than 300\,ms at high $\rho$ and $\mathrm{OF}$. This indicates superior computational efficiency on our implementation, making it highly suitable for real-time applications.

\begin{table}[t]
\centering
\caption{Average execution time of RSoD, ITER-SIS, and LASSO-B$^2$R$^2$ for $\rho$ and $\OF$ with Input SNR at $19.40$ dB.}
\renewcommand{\arraystretch}{1.2}
\begin{tabular}{c|c|c|c|c}
\toprule
\hline
\multirow{2}{*}{$\rho$} & \multirow{2}{*}{$\OF$} & \multicolumn{3}{c}{Average Algorithm Run Time (ms)} \\
\cline{3-5} 
 & & RSoD 
& ITER-SIS~\cite{Guo2023_ICASSP_ITERSIS} 
& LASSO-B$^2$R$^2$~\cite{Shah2024_LASSO_B2R2} \\ \hline
3 & 5 & 1.48 & 14.41 & 4.25 \\ \hline
8 & 15 & 1.54 & 22.71 & 48.76 \\ \hline
15 & 30 & 1.55 & 53.31 & 323.85 \\ 
\hline
\bottomrule
\end{tabular}
\label{tab:alg_comparison_time_sim}
\end{table}
\textbf{Simulation 3: Oversampling Requirements under variant $\rho$, $\rho_\eta$ and $\nu$.}
In this simulation, we generate the test signals in the same way as in Simulation 2 and compare empirical results with our theoretical analysis.  

\begin{figure*}[t]
    \centering
    \subfloat[OF vs.\ $\rho$]{
        \includegraphics[width=0.31\textwidth]{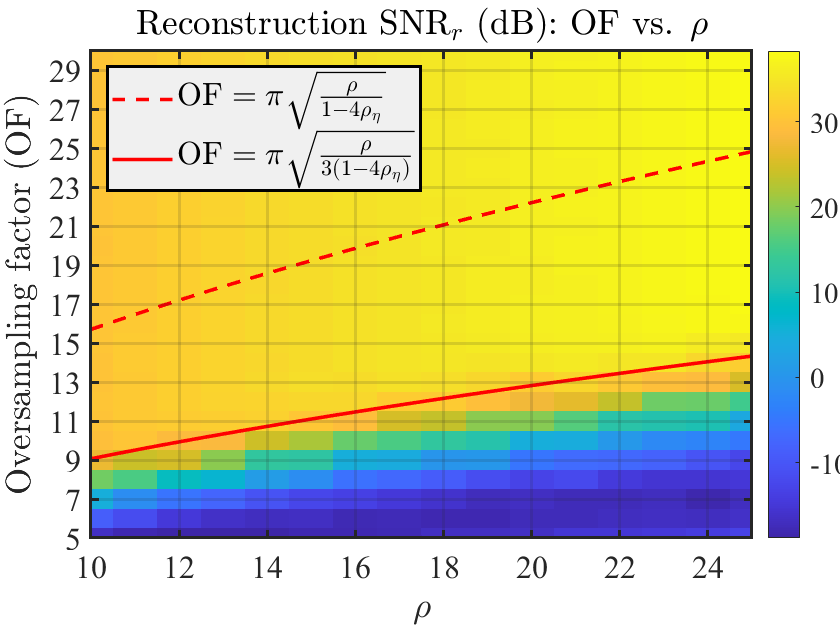}
        \label{fig:OFvsrho}}
    %\hfill
    \hspace{-0.2cm}
    \subfloat[OF vs.\ $\rho_{\eta}$]{
        \includegraphics[width=0.31\textwidth]{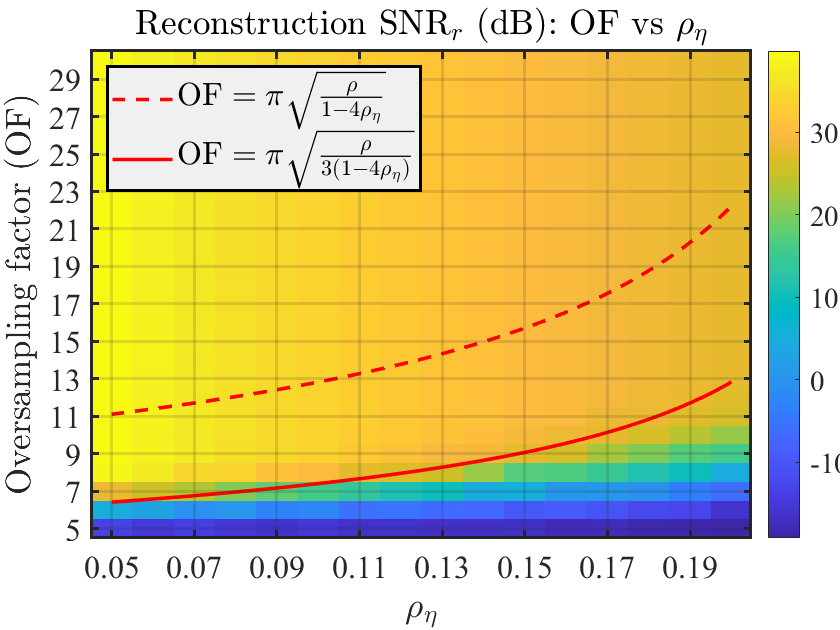}
        \label{fig:OFvsrhoeta}}
    %\hfill
    \hspace{-0.2cm}
    \subfloat[OF vs.\ $\nu$]{
        \includegraphics[width=0.31\textwidth]{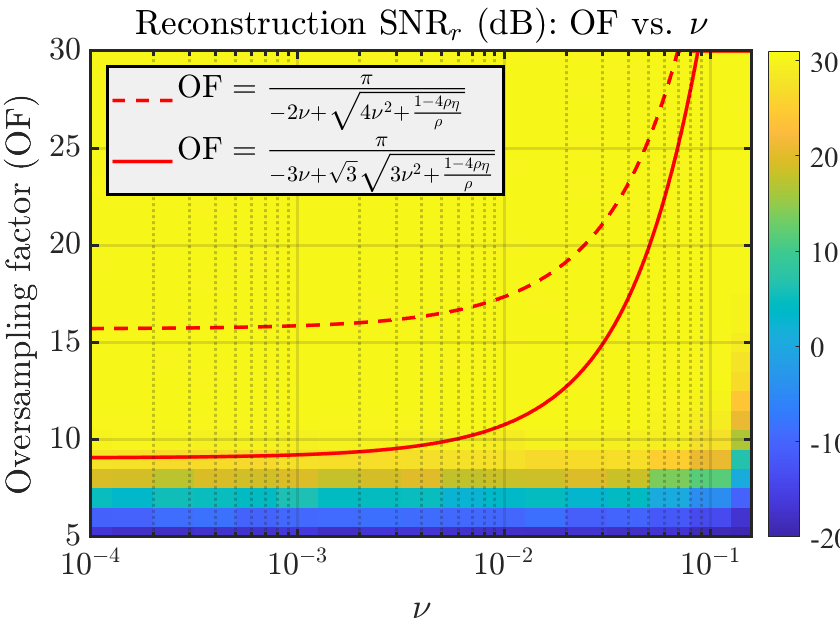}
        \label{fig:OFvsnu}}
    \caption{Phase transition diagrams comparing empirical reconstruction SNR with theoretical predictions: 
    (a) OF versus amplitude scaling $\rho$, 
    (b)  OF versus normalized noise level
    and $\rho_\eta$ (c) OF versus jitter parameter $\nu$.}
    \label{fig:phase_transition}
\end{figure*}

\subsubsection{OF vs. $\rho$}
We evaluate the reconstruction performance across $\rho \in [10,25]$ and $\OF \in [5,30]$. Modulo samples are corrupted by measurement noise uniformly distributed as $\mathcal{U}[-0.15\lambda,\,0.15\lambda)$, corresponding to $\rho_\eta=0.15$. The resulting $\text{SNR}_r$ for each $(\rho,\OF)$ pair is averaged over 500 trials.

Fig.~\ref{fig:phase_transition}(a) presents the empirical phase transition diagram (SNR vs. $\OF$ and $\rho$) together with the theoretical RSoD thresholds. The predicted oversampling requirements, both the generic bound (Eq.~\eqref{eq:RSoD_OF}) and the refined sinc-based bound Eq.~\eqref{eq:OFRSOD_sinc} are plotted for comparison. The close alignment of the empirical results with the sinc-specific bound confirms the tightness of the RSoD analysis in the presence of bounded noise. Note that 3-bit quantization yields a maximum noise level of $\rho_\eta \leq 0.125$, which is lower than the $0.15$ used here. This suggests that RSoD is inherently robust to the noise introduced by quantization.

\subsubsection{$\OF$ vs. $\rho_\eta$}
Using the input signals defined in~\eqref{eq:sincSim} with fixed $\rho=10$, we vary $\rho_\eta$ from $0.05$ to $0.20$ and $\OF$ from $5$ to $30$, the reported SNR$_r$ is averaged over $500$ independent realizations. Fig.~\ref{fig:phase_transition}(b) shows the average reconstruction SNR$_r$ versus $(\rho_\eta,\OF)$. A clear phase transition is observed: for small $\rho_\eta$, reliable recovery is achieved with modest oversampling ($\OF \approx 6$--$8$), whereas larger $\rho_\eta$ up to $0.2$ requires substantially higher $\OF$. The empirical transition closely follows the sinc-specific prediction in~\eqref{eq:OFRSOD_sinc}.
%\vspace{-0.2cm}
\subsubsection{$\OF$ vs. $\nu$}
We evaluate the oversampling factor ($\OF$) against the jitter parameter $\nu$ using the input signals from~\eqref{eq:sincSim} with $\rho=10$ and $\rho_\eta=0.15$. While practical jitter values $\nu$ are typically on the order of $10^{-3}$ to $10^{-4}$, we vary $\nu$ up to $10^{-1}$ to stress test the system. As shown in Fig.~\ref{fig:phase_transition}(c), the empirical $\OF$ remains nearly constant even for large $\nu$ ($10^{-2}$--$10^{-1}$). This is in contrast to theoretical bounds, which predict a sharp increase in $\OF$. The result demonstrates that RSoD is highly robust to jitter, enabling the overall pipeline to function far beyond the $10^{-3}$--$10^{-4}$ range. This robustness confirms the potential for future work combining modulo ADCs with compressed sensing by applying CS techniques directly to non-uniform samples recovered by RSoD.

\section{Hardware Experiments}\label{sec:exp}

\begin{figure*}[t]
    \centering
    \includegraphics[width=1\linewidth]{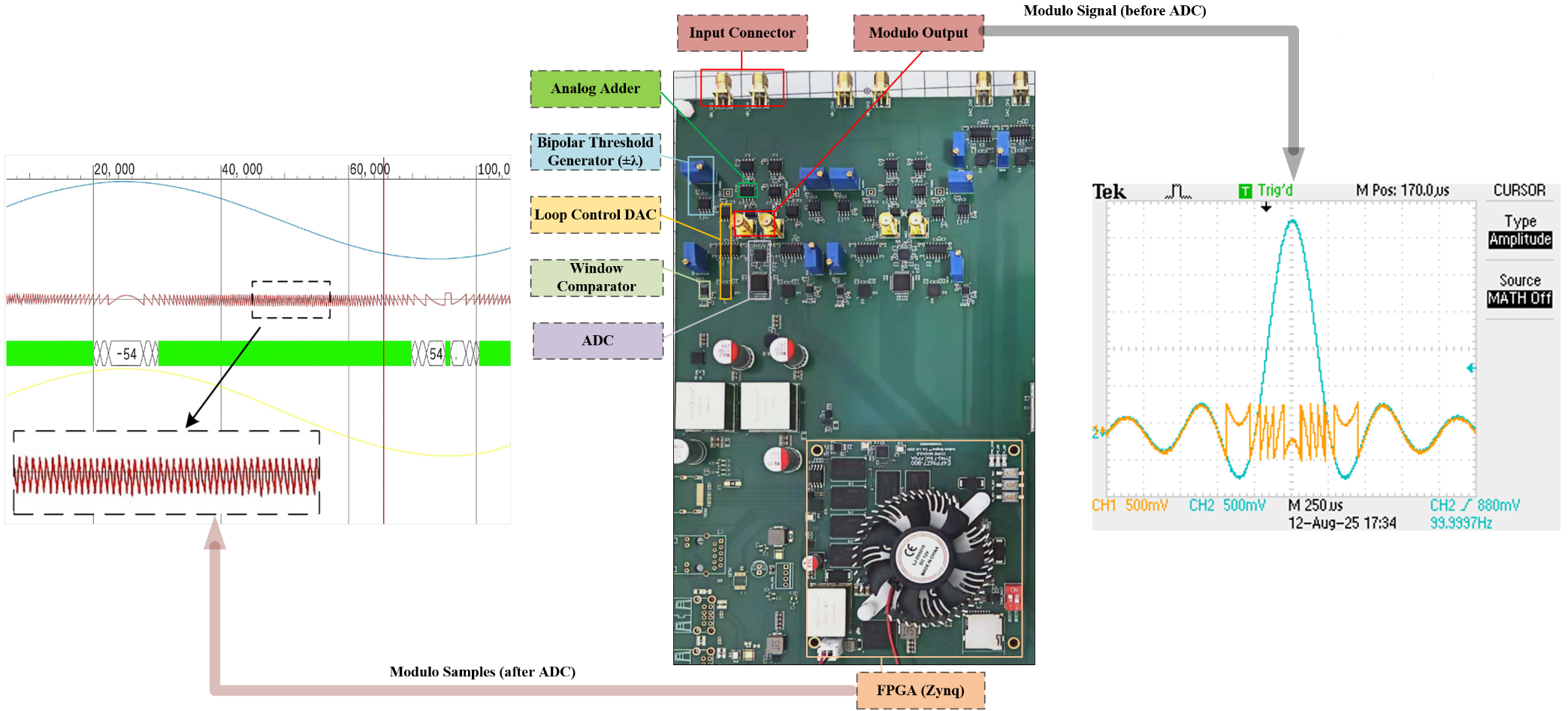}
\caption{FPGA-based modulo ADC prototype and measurement results. 
The middle panel shows the fabricated hardware board with labelled functional blocks. 
The left panel displays FPGA-captured waveforms after ADC conversion (blue: original input, red: folded signal, green: folding counts). 
The right panel shows oscilloscope traces measured directly (blue: analog input, yellow: folded modulo output).}
    \label{fig:hardwarePro}%\vspace{-0.3cm}
\end{figure*}

The previous simulations confirm reliable RSoD performance at high SNR. In the \si{\kilo\hertz} range, prior hardware demonstrations~\cite{Bhandari2022TSP_FP,mulleti_hardware_2023,Zhu_TIM_2024} achieved operation up to \SI{15}{\kilo\hertz} with $\rho < 12$, but distortions in the folded waveform limited achievable performance. To obtain high-quality modulo samples for validation, we developed an FPGA-based prototype that leverages high clock rates and digital control (Fig.~\ref{fig:hardwarePro}). While the detailed hardware design will be reported separately, here we use this platform to experimentally validate RSoD across the \si{\kilo\hertz} range with amplitude expansion up to $\rho=108$.

% Fig.~\ref{fig:hardwarePro} depicts the fabricated prototype board employed in our experiments, serving as the hardware platform for validating the theoretical analysis and simulations presented in the previous sections. 
Following~\cite{Bhandari2022TSP_FP,Guo2023_ICASSP_ITERSIS,zhu_60_10Hz}, the analog input (ground truth) and modulo output are captured by an 8-bit Tektronix TDS\,1012C-EDU oscilloscope. RSoD and baseline algorithms are applied offline in \textsc{matlab}. 
The on-board ADC is bypassed to (\emph{i}) adjust oversampling via software decimation and (\emph{ii}) avoid ADC-induced gains that affect the calculation of reconstruction SNR$_r$.

\begin{figure}
    \centering
    \includegraphics[width=1\linewidth]{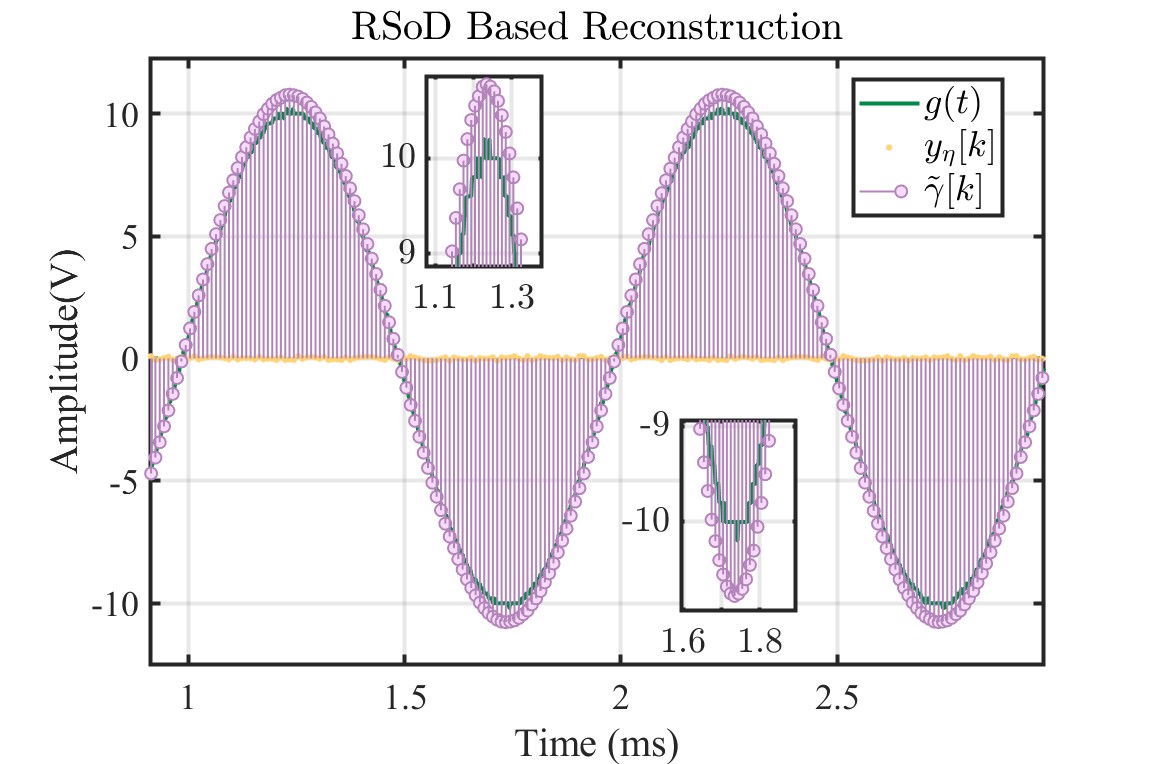}
    \caption{Reconstruction of a \SI{1}{\kilo\hertz} sinusoid ($\lambda=\SI{0.1}{\volt}$, $\rho=108$, $\OF=50$ and $b=3$). The original signal $g(t)$ (green) is folded and sampled to yield $y_\eta[k]$ (yellow), and accurately recovered as $\tilde{\gamma}[k]$ (pink) using RSoD.}\vspace{-0.3cm}
    \label{fig:rho100_sine}
\end{figure}

\begin{table}[t]
\centering
\caption{SINAD and ENOB comparison for a \SI{1}{kHz} sinusoid ($\rho=108$, $\lambda=\SI{0.1}{\volt}$). 
The conventional ADC full-scale matches the input peak, while the proposed system uses modulo sampling with RSoD recovery.}
\label{tab:sinad_enob_gain}
\renewcommand{\arraystretch}{1.3}
\begin{tabular}{c|cc|cc}
\toprule
\hline
\multirow{2}{*}{Bits} & \multicolumn{2}{c|}{Conventional ADC} & \multicolumn{2}{c}{Proposed System}\\
\cline{2-5}
 & SINAD (\si\decibel)& ENOB (bits) & SINAD (\si\decibel)& ENOB (bits) \\
\hline
3 & 19.66   & 2.97 & 57.14 & 9.20 \\
\hline
4 & 25.59  & 3.96 & 58.48 & 9.42 \\
\hline
5 & 31.19  & 4.89  & 59.17  & 9.54 \\
\hline
6 & 36.66  & 5.80  & 59.31 & 9.56 \\
\hline
7 & 38.70 & 6.14  & 59.47 & 9.59 \\
\hline
8 & 39.88 &  6.33  & 59.47 & 9.59 \\
\hline
\bottomrule
\end{tabular}
\end{table}

\textbf{Experiment 1: Comparison with a Conventional ADC.} 
Although our theory treats finite-energy bandlimited signals, we begin hardware validation with a single cosine input. This choice is deliberate: for $x(t)=\cos(\Omega t)$ one has $\|x^{(2)}(t)\|_\infty=\Omega^2\|x\|_\infty$, i.e., the cosine saturates the classical Bernstein inequality for second-order derivatives and thus serves as a worst-case test for RSoD. It also enables a straightforward computation of SINAD in \textsc{matlab}.

The proposed modulo-ADC with RSoD recovery is compared with a conventional ADC for a \SI{1}{\kilo\hertz} sinusoid ($\lambda=\SI{0.1}{\volt}$, $\rho=108$), with the full-scale set to the input peak. The signal is sampled at $f_s=\SI{100}{\kilo\hertz}$ (i.e., $\OF=50$) and recovered using RSoD. Performance is evaluated using a $b$-bit uniform quantizer ($3\leq b\leq 8$) in terms of SINAD and ENOB. For comparison, we also include the signal directly captured by the oscilloscope. 

Table~\ref{tab:sinad_enob_gain} confirms Corollary~\ref{cor:conventional}. At low resolution ($b=3$), the modulo ADC achieves a gain of 37.5\,dB in SINAD and 6.2\,bits in ENOB, close to the theoretical values of 40.7\,dB and 6.8\,bits. As the resolution increases ($b=4$–5), the gain decreases because quantization noise becomes less dominant. At higher resolutions ($b=6$–8), the gain stabilizes at around 21\,dB and 3.5\,bits, limited by front-end noise and folding distortions. Fig.~\ref{fig:rho100_sine} illustrates the reconstruction for $b=3$. Due to the oscilloscope’s limited dynamic range, the input signal exhibits distortion when $|g(t)|>10$, whereas modulo sampling with RSoD produces a smoother recovery. These observations are consistent with those reported in~\cite{zhu_60_10Hz} for a 10\,Hz sinusoid.

For context, the hardware experiment in~\cite{Bhandari2022TSP_FP} reports a 50\,Hz signal sampled at 
$T=108.488\,\mu\text{s}$ ($f_s\!\approx\!9.218$\,kHz), corresponding to an oversampling factor 
$\text{OF}\!\approx\!92.18$ for $\rho=24$. As the two setups differ in hardware, thresholds, and noise and distortions, this comparison is indicative and illustrates the capability of modulo ADCs rather than providing a direct benchmark. We note that our \SI{1}{\kilo\hertz} test is a second-order stress case (since the bound scales with $\Omega^2$), yet reliable reconstruction is obtained at $\text{OF}=50$ for $\rho=108$, demonstrating the robustness of the modulo-ADC with RSoD recovery.

\begin{table*}[t]
\centering
\caption{Performance comparison of RSoD, ITER-SIS~\cite{Guo2023_ICASSP_ITERSIS}, and LASSO-B$^2$R$^2$~\cite{Shah2024_LASSO_B2R2} for a sinc input across multiple $(B,\rho,\lambda)$ configurations. Metrics: OF (dark red = smallest), SNR$_r$, and PSNR (bold = highest).}
\renewcommand{\arraystretch}{1.3}
\begin{tabular}{c|c|c|ccc|ccc|ccc}
\toprule
\hline
\multicolumn{2}{c|}{\begin{tabular}[c]{@{}c@{}}Input Signal \\ Parameters (Sinc)\end{tabular}}
& \multicolumn{1}{c|}{\begin{tabular}[c]{@{}c@{}}Modulo ADC\\ Threshold \end{tabular}} 
& \multicolumn{3}{c|}{RSoD} 
& \multicolumn{3}{c|}{ITER-SIS~\cite{Guo2023_ICASSP_ITERSIS}} 
& \multicolumn{3}{c}{LASSO-B$^2$R$^2$~\cite{Shah2024_LASSO_B2R2}} \\
\hline 
$B$ (\si{\kilo\hertz})& $\rho$ & $\lambda\,(\si{\volt}) $  
& $\mathrm{OF}$ & SNR$_{r}$(\si\decibel)& PSNR (\si\decibel)& $\mathrm{OF}$  & SNR$_{r}$(\si\decibel)& PSNR (\si\decibel)& $\mathrm{OF}$  & SNR$_{r}$(\si\decibel)& PSNR (\si\decibel)\\ \hline

2   & 102.6 & 0.10  & \textcolor{deepred}{35.71} & \textbf{19.43} & \textbf{27.32} & 50.00 & 2.05 & 11.45 & 35.71  & 0.24 & 10.09 \\ \hline

20  & 16.48  & 0.48  & \textcolor{deepred}{10.42} & \textbf{22.03} & \textbf{37.35} & 20.83 & 11.62 & 26.86 & 31.25  & 11.88 & 24.27 \\ \hline

50  & 11.31  & 0.48 & \textcolor{deepred}{10.00} & \textbf{30.52} & \textbf{42.48} & 25.00 & 19.45 & 31.16 & 16.67  & 28.44 & 41.03 \\ \hline

180 &  3.71 & 0.36  & \textcolor{deepred}{3.66} & 33.12 & 42.36 & 3.97 & 31.86 & 42.13 &  3.86 & \textbf{33.89} & \textbf{44.02}  \\ \hline

400 &  3.32 & 0.36  &3.68 & 29.11 & 39.88 & 3.47 & \textbf{30.70} & \textbf{41.61} &  \textcolor{deepred}{3.38} & 26.65 & 37.65 \\
\hline
\bottomrule
\end{tabular}
\label{tab:alg_comparison_sinc}
\end{table*}

\begin{table*}[t]
\centering
\caption{Performance comparison of RSoD, ITER-SIS~\cite{Guo2023_ICASSP_ITERSIS}, and LASSO-B$^2$R$^2$~\cite{Shah2024_LASSO_B2R2} for general bandlimited inputs across multiple $(B,\rho,\lambda)$ configurations. Metrics: OF (dark red = smallest), SNR$_r$, and PSNR (bold = highest).}
\renewcommand{\arraystretch}{1.3}
\begin{tabular}{c|c|c|ccc|ccc|ccc}
\toprule
\hline
\multicolumn{2}{c|}{\begin{tabular}[c]{@{}c@{}}General Bandlimited \\ Signal \end{tabular}}
& \multicolumn{1}{c|}{\begin{tabular}[c]{@{}c@{}}Modulo ADC\\ Threshold\end{tabular}} 
& \multicolumn{3}{c|}{RSoD} 
& \multicolumn{3}{c|}{ITER-SIS~\cite{Guo2023_ICASSP_ITERSIS}} 
& \multicolumn{3}{c}{LASSO-B$^2$R$^2$~\cite{Shah2024_LASSO_B2R2}} \\
\hline 
$B$ (\si{\kilo\hertz})& $\rho$ & $\lambda\,(\si{\volt}) $ 
& $\mathrm{OF}$
& SNR$_{r}$(\si\decibel)& PSNR (\si\decibel)& $\mathrm{OF}$  & SNR$_{r}$(\si\decibel)& PSNR (\si\decibel)& $\mathrm{OF}$  & SNR$_{r}$(\si\decibel)& PSNR (\si\decibel)\\ \hline

1   & 20.50 & 0.10 & \textcolor{deepred}{11.36} & 14.37 & 25.87 & 41.67 & 8.05 & 21.27 & 41.67  & \textbf{22.63} & \textbf{34.29} \\ \hline

10   & 7.15 & 0.48 & \textcolor{deepred}{5.00} & \textbf{31.73} & \textbf{42.79} & 10.00 & 29.49 & 41.71 & 8.33  & 29.64 & 40.72 \\ \hline

20   & 7.23 & 0.48 & \textcolor{deepred}{6.25} & 27.44 & 40.58 & 10.41 & \textbf{29.75} & 41.09 & 8.93 & 28.88 & \textbf{41.12} \\ \hline

20  & 17.28  & 0.36 & \textcolor{deepred}{8.93} & \textbf{28.51} & \textbf{38.99} & 31.25 & 9.83 & 21.21 & 31.25  & 12.65 & 24.30 \\ \hline

100 & 5.92  & 0.36 & \textcolor{deepred}{4.55} & 28.75 & 40.05 & 6.25 & \textbf{30.21} & \textbf{40.67} & 4.94 & 26.62 & 37.73 \\
\hline
\bottomrule
\end{tabular}
\label{tab:alg_comparison_bw}
\end{table*}

\begin{figure}[t]
    \centering
    \includegraphics[width=0.95\linewidth]{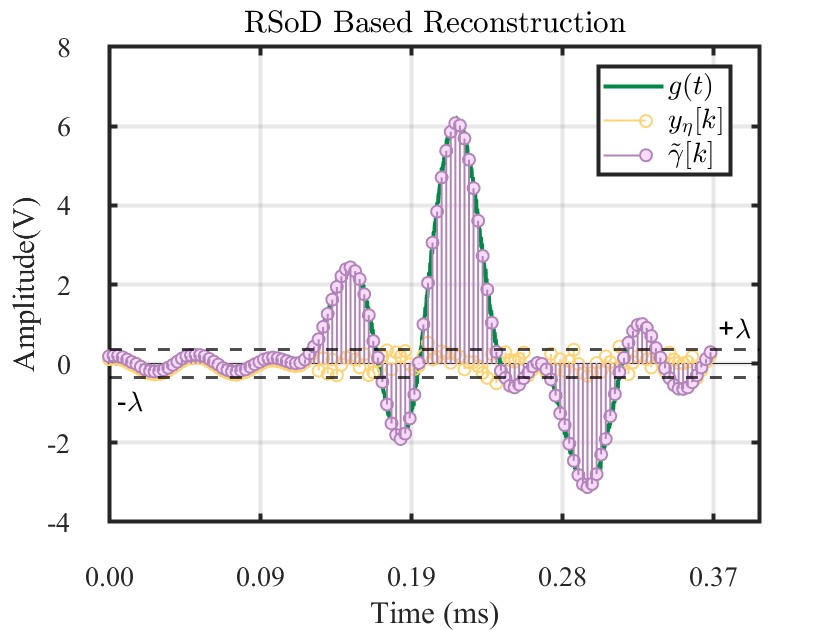}
\caption{Reconstruction of a general bandlimited input ($B=\SI{20}{\kilo\hertz}$, $\rho = 17.28$, $\OF=8.93$ and $\lambda = \SI{0.36}{\volt}$). 
The original signal $g(t)$ (green) is folded and sampled to yield $y_\eta[k]$ (yellow), and recovered as $\tilde{\gamma}[k]$ (pink) using RSoD.}
    \label{fig:reus_general_bandlimited}
\end{figure}

\textbf{Experiment 2: Comparison with Existing Recovery Algorithms.}
We benchmark against \textbf{ITER-SIS}~\cite{Guo2023_ICASSP_ITERSIS} and \textbf{LASSO–B$^2$R$^2$}~\cite{Shah2024_LASSO_B2R2}.
A signal generator (Tektronix AFG1022) produces sinc functions and general bandlimited inputs (configured via software) across varying bandwidth $B$, scaling factor $\rho$, and ADC threshold $\lambda$ for experimental validation.

\subsubsection{OF and Reconstruction SNR}
Performance is first evaluated in terms of required OF, reconstruction $\text{SNR}_{r}$~\eqref{eq:rrse_def}, and the peak signal-to-noise ratio (PSNR)
\[
\mathrm{PSNR} = 10 \log_{10}\!\left(
\frac{\|g\|_\infty^2}{\frac{1}{L}\sum_{k=1}^{L}\big(\gamma[k]-\tilde{\gamma}[k]\big)^2}
\right),
\]
where $L$ is the signal length. 

As seen from Tables~\ref{tab:alg_comparison_sinc} and~\ref{tab:alg_comparison_bw}, for large dynamic-range expansion ($\rho \geq 5$), RSoD achieves competitive or higher reconstruction quality while requiring substantially lower oversampling factors. When $\rho$ decreases ($\rho \approx 3$–4), the performance of all methods converges. For general bandlimited signals, RSoD typically operates with $\text{OF} \approx 4.5$–$11.4$, a factor of 2–4 lower than ITER-SIS and LASSO, while achieving comparable or better SNR$_r$ and PSNR. We note that in one case (Table~\ref{tab:alg_comparison_bw}, $B=1$ kHz and $\rho=20.50$), LASSO-B$^2$R$^2$ achieves higher SNR$_r$, but only at a much larger $\OF$. This illustrates the distinct rate–robustness trade-offs across the algorithms: RSoD emphasizes sampling efficiency, LASSO favors stability under noise at the cost of higher $\OF$, and ITER-SIS provides good performance for small $\rho$ but degrades at larger $\rho$ due to numerical instabilities.

\subsubsection{OF Bounds: Theory vs. Experiment}
Table~\ref{tab:alg_comparison_of_sinc} validates the theoretical OF bounds for RSoD across various bandwidth and dynamic range conditions using 8-bit quantization ($\rho_{\eta}=2^{-8}$). The experimental OF consistently falls between the generic bound of Eq.~\eqref{eq:RSoD_OF} and the sinc-specific bound of Eq.~\eqref{eq:OFRSOD_sinc}, confirming the theoretical framework. Since the low-pass filtered signals used in experiments approximate brick-wall characteristics, the measured OF values closely track the tighter sinc bound, demonstrating the $\sqrt{3}$ improvement over the generic estimate. This agreement validates both the noise-aware sampling theory and the practical effectiveness of the RSoD approach across the tested parameter space.

\begin{table}[t]
\centering
\caption{Comparison of theoretical and experimental oversampling factors for RSoD recovery.}
\renewcommand{\arraystretch}{1.2}
\begin{tabular}{c|c|c|c|c}
\toprule
\hline
\multirow{2}{*}{\begin{tabular}[c]{@{}c@{}}$B$ \\ (kHz)\end{tabular}} & \multirow{2}{*}{$\rho$} & \multicolumn{3}{c}{OF} \\
\cline{3-5}
& & \begin{tabular}[c]{@{}c@{}}Eq.~\eqref{eq:RSoD_OF}\end{tabular} & \begin{tabular}[c]{@{}c@{}}Eq.~\eqref{eq:OFRSOD_sinc}\end{tabular} & Experiment \\ 
\hline
1   & 20.50 & 14.22 & 8.21 & 11.36\\\hline 
10  & 7.15 & 8.40 & 4.85 & 5.00 \\ \hline 
20  & 7.20 & 8.43  & 4.87 & 6.25 \\\hline  
20 &  17.28 & 13.06  &7.54 & 8.93 \\\hline  
100 &  5.92 & 7.64 & 4.41 & 4.55 \\\hline
\bottomrule
\end{tabular}
\label{tab:alg_comparison_of_sinc}
\end{table}

\textbf{Experiment 3: Robustness to Noisy Input Signals}
We evaluate RSoD's robustness to additive Gaussian noise in the input signal $x(t)$ with $B{=}\SI{50}{\kilo\hertz}$, $\rho{=}11.72$, and $\lambda{=}\SI{0.36}{\volt}$. The signal is passed through a second-order Butterworth filter before modulo sampling to produce $g(t)$. We test $\mathrm{OF}{=}8.33$ and $10.00$ with input SNR$_x \in \{\SI{5}{\decibel}, \SI{10}{\decibel}, \SI{15}{\decibel}, \SI{20}{\decibel}\}$. 

Table~\ref{tab:snr_results} summarizes the results. At low input SNR (\SIrange{5}{10}{\decibel}), RSoD preserves reconstruction SNR$_r$ close to the input level. At higher input SNR (\SIrange{15}{20}{\decibel}), RSoD achieves improved reconstruction with SNR$_r$ exceeding \SI{20}{\decibel}, demonstrating robust noise handling across the tested range.

\begin{table}[t]
\centering
\caption{SNR$_r$ under additive Gaussian noise for a bandlimited input ($B=\SI{50}{\kilo\hertz}$, $\rho=11.72$, $\lambda=\SI{0.36}{\volt}$) at $\mathrm{OF}= 8.33$ and $10.00$.}
\renewcommand{\arraystretch}{1.3}
\begin{tabular}{c|c|c|c|c}
\toprule
\hline
OF      & \multicolumn{2}{c|}{8.33} & \multicolumn{2}{c}{10.00} \\
\hline
SNR$_x$ (dB) & 5   & 10   & 15   & 20   \\
\hline
SNR$_r$ (dB) & 5.30 & 10.51 & 20.73 & 23.55 \\
\hline
\bottomrule
\end{tabular}
\label{tab:snr_results}
\end{table}

\textbf{Experiment 4: Reconstruction from Non-uniform Samples.}
To verify Theorem~\ref{thm:noisy_N2_jitter}, we consider non-uniform modulo samples of an input signal with parameters 
$(B=\SI{100}{\kilo\hertz},\, \rho = 5.92,\, \lambda = \SI{0.36}{\volt})$. 
Random sampling jitters with deviation $\nu=0.09$ were introduced into the oscilloscope-captured time instants. 
The RSoD algorithm was applied to unwrap the modulo samples, and the continuous-time signal was subsequently recovered using a NUDFT/NUFFT-based reconstruction method~\cite{dutt1993fast,potts2001fast,greengard2004accelerating}.

\begin{figure}[t]
    \centering
    \subfloat[Sampling-interval distribution]{\includegraphics[width=0.8\linewidth]{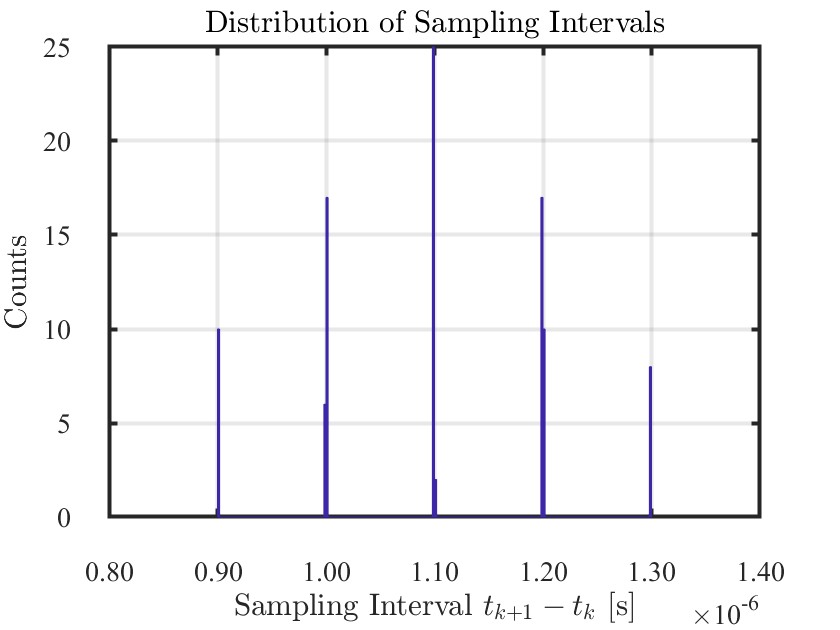}
        \label{fig:jitter_hist}}\\
    %\hfill
    \subfloat[Recovered vs.\ oscilloscope captured original signal]{\includegraphics[width=0.8\linewidth]{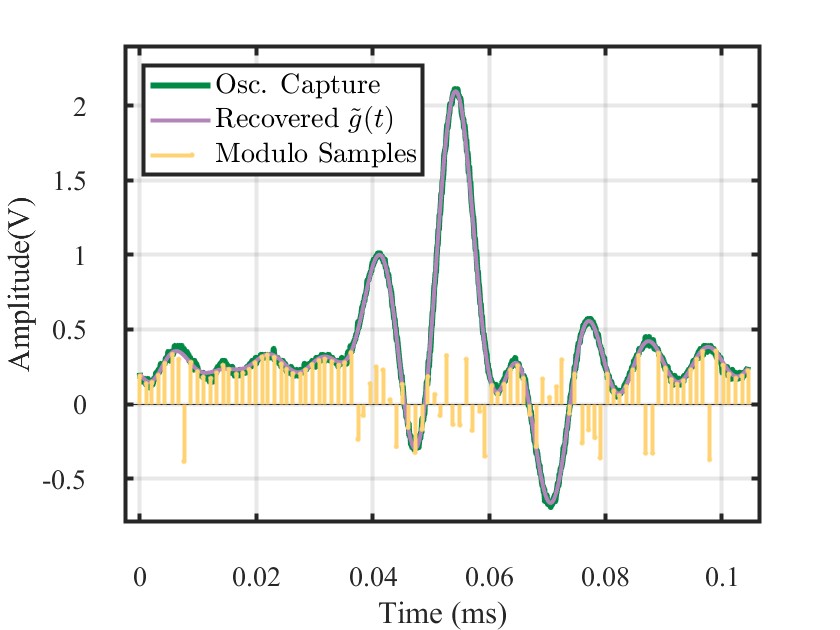}
        \label{fig:recovered_signal}}
             \caption{Signal recovery from jittered non-uniform modulo samples ($B = \SI{100}{\kilo\hertz}$, $\rho = 5.92$, $\lambda = \SI{0.36}{\volt}$, $\text{OF} = 4.55$). (a) Sampling interval histogram ($\nu=0.09$); (b) Recovered vs. original signal and modulo samples.}\label{fig:nu_modulo_recovery}  
\end{figure}

Fig.~\ref{fig:nu_modulo_recovery}(a) shows the histogram of the sampling-interval distribution, which confirms the presence of random jitter around the nominal period. 
Fig.~\ref{fig:nu_modulo_recovery}(b) compares the reconstructed signal with the waveform captured directly by the oscilloscope. 
The results demonstrate that even with non-uniform samples, the proposed RSoD framework is able to faithfully recover the underlying signal. 
In fact, the recovered waveform is noticeably smoother and less noisy than the raw oscilloscope trace, underscoring the robustness of the approach. 
To the best of our knowledge, this is the first work to investigate recovery from non-uniform modulo samples.

\section{Conclusion and Future Work}\label{sec:con}
\subsubsection*{Summary of the paper's contributions} This work revisited modulo (unlimited) sampling from both a theoretical and practical perspective. 
We analysed the difference-based recovery framework and established refined conditions on the sufficient oversampling factor, showing that classical bounds can be substantially tightened. 
The RSoD scheme was investigated in detail, where its balance between noise robustness and sampling efficiency makes it particularly suitable for implementation. 
Beyond the idealized uniform setting, we derived explicit bounds for the case of non-uniform samples, demonstrating that recovery remains stable even under bounded timing jitter, an aspect not previously addressed in the unlimited sensing literature. 
Experimental evaluation on an FPGA prototype confirmed that the theory translates into practice, achieving accurate recovery across the kilohertz range with more than 100 times of magnitude amplitude expansion. 

\subsubsection*{Limitations and Future Work}
While the revised second-order difference (RSoD) method provides sharp
theoretical guarantees and strong performance at moderate oversampling
rates, it also has limitations in practice. In our FPGA-based prototype,
the digital logic operates reliably at 200~MHz, ensuring high-quality
modulo samples in the kHz regime. However, for higher-frequency inputs
($>450$~kHz), non-idealities in the analog front-end, including
hysteresis and slew-rate limitations, introduce impulsive distortions
in the folded waveform. These artifacts resemble the impulsive noise
patterns reported in Fourier--Prony recovery~\cite{Bhandari2022TSP_FP}
and effectively violate the bounded-noise assumption
($\rho_\eta < 1/4$). Under such conditions, RSoD becomes unstable since
the anti-differencing step amplifies impulsive errors, leading to error
propagation in the reconstruction. Addressing this issue will require
both improved hardware design (to extend stable operation at higher
bandwidths) and algorithmic advances, for example by detecting and
suppressing impulsive noise prior to RSoD unwrapping. These directions
are left for future work.

\section*{Acknowledgments}
 The authors are grateful to Prof. Ayush Bhandari, Dr. Ruiming Guo, and Mr. Yuliang Zhu (Imperial College London) for insightful discussions and their prior contributions to modulo ADCs. The authors would also like to thank Mr. Thiruganesan Dhanendraraj, Mr. Michael Lateo, and Mr. William Schkzamian (Brunel University of London) for their guidance and assistance during the hardware development process.

\appendices
\section{Divided Differences and Forward Differences}\label{sec:Apdif}

For completeness, we recall several standard definitions and results on divided differences~\cite{deBoor2005}, which are used in the proof of Lemma~\ref{lem:FD-bound}. We first state the general definition of divided differences for arbitrary nodes, followed by the special case of equally spaced nodes, i.e., forward differences. These notions provide the link between discrete difference operators and continuous derivatives.

\begin{deft}[Divided Difference]
Let $g(t)$ be a function, and let 
$t_0<t_1<\cdots<t_L \in \mathbb{R}$ be pairwise distinct nodes. 
The divided differences of $g$ are defined recursively as
\begin{gather}
    [\,g(t_0)\,] \triangleq g(t_0), \\[6pt]
    \begin{split}
    &\,[\,g(t_0),g(t_1),\ldots,g(t_k)\,] 
    \\\triangleq &\frac{[\,g(t_1),\ldots,g(t_k)\,] 
          - [\,g(t_0),\ldots,g(t_{k-1})\,]}
          {\,t_k - t_0\,}.
          \end{split}
\end{gather}
\end{deft}

\begin{deft}[Forward Difference]
If the nodes $t_k$ ($0\le k\le L$) are equally spaced with step size $T>0$, i.e.,
\[
t_k = t_0 + k T, \quad k=0,1,\ldots,N,
\]
then the divided differences reduce to the usual forward differences, defined recursively by
\begin{align}
    \Delta^{(0)} g(t_k) &\triangleq g(t_k),\\ 
    \Delta^{(j)} g(t_k) &\triangleq \Delta^{(j-1)} g(t_{k+1}) 
        - \Delta^{(j-1)} g(t_k)
\end{align}
for $k=0,1,\ldots,L$ and $j=1,2,\ldots,N,\; k=0,1,\ldots,N-j $
\end{deft}

The following mean value theorem shows that an $N$-th order divided difference equals the $N$-th derivative of the function at some intermediate point, scaled by $1/N!$. 

\begin{theorem}[Mean Value Theorem for Divided Differences]
\label{thm:MVT-divdiff}
Let $t_0<t_1<\cdots<t_L$ be distinct nodes, and suppose 
$g$ is $N$-times continuously differentiable on the interval 
$(t_0,t_L)$. Then for each $k=0,1,\ldots,L-N$, there exists a point 
$\xi_k\in(t_k,t_{k+N})$ such that~\cite{deBoor2005}
\begin{equation}
    [\,g(t_k),g(t_{k+1}),\ldots,g(t_{k+N})\,]
    \;=\; \frac{g^{(N)}(\xi_k)}{N!}.
\end{equation}
\end{theorem}

In the case of equally spaced nodes, the corollary below gives the mean value representation of forward differences, a result that is essential for establishing Lemma~\ref{lem:FD-bound}.

\begin{corollary}[Uniform-Grid (Forward) Mean Value Form]
\label{cor:forward-MVT}
Suppose $g$ is $N$-times continuously differentiable on $(t_k,t_{k+N})$.
Then for each $k=0,1,\ldots,L-N$ there exists
$\xi_k\in(t_k,t_{k+N})$ such that~\cite{deBoor2005}
\begin{equation}
    \Delta^{(N)} g(t_k) \;=\; T^{N}\, g^{(N)}(\xi_k).
\end{equation}
\end{corollary}

\begin{IEEEproof}
For equally spaced nodes one has the identity
\[
[\,g(t_k),\ldots,g(t_{k+N})\,] \;=\; \frac{\Delta^{(N)} g(t_k)}{N!\,T^{N}} .
\]
By Theorem~\ref{thm:MVT-divdiff}, there exists 
$\xi_k\in(t_k,t_{k+N})$ such that
\[
[\,g(t_k),\ldots,g(t_{k+N})\,] \;=\; \frac{g^{(N)}(\xi_k)}{N!}.
\]
Combining the two relations gives the desired result.
\end{IEEEproof}

\section{ Proof of Theorem~\ref{thm:noisy}}\label{Asec:noisy_bound}
\begin{IEEEproof}
We first establish that condition~\eqref{eq:noisy_reccond} ensures
\begin{equation}
\mathcal{M}_\lambda(\Delta^N y_\eta)-\Delta^N y_\eta=\Delta^N\varepsilon_{\gamma}.
\end{equation}
By Lemma~\ref{lem:FD-bound} and the triangle inequality, we can bound $|\Delta^N \gamma+\Delta^N \eta|$:
\begin{align}
|\Delta^N \gamma+\Delta^N \eta|
&\leq |\Delta^N \gamma|+|\Delta^N \eta| \nonumber\\
&\leq (T\Omega)^{N} \|g\|_{\infty}+\|\eta\|_{\infty} \cdot 2^{N} < \lambda,
\end{align}
which implies
\begin{equation}\label{eq:boundgamma_eta}
\mathcal{M}_\lambda (\Delta^N \gamma+\Delta^N \eta)=\Delta^N \gamma+\Delta^N \eta.
\end{equation}

Following the approach in~\cite{bhandari_unlimited_2021}, we have
\begin{align}
&\mathcal{M}_{\lambda}(\Delta^{N} y_{\eta}) - \Delta^{N} y_{\eta} \nonumber\\
&= \mathcal{M}_{\lambda}(\Delta^{N} y+\Delta^{N}\eta)- \Delta^{N} y - \Delta^{N} \eta \nonumber\\
&= \mathcal{M}_{\lambda}(\mathcal{M}_{\lambda}(\Delta^{N} y)+\Delta^{N} \eta) - \Delta^{N} y - \Delta^{N} \eta \nonumber\\
&= \mathcal{M}_{\lambda}(\Delta^{N} \gamma + \Delta^{N} \eta) - \Delta^{N} y - \Delta^{N} \eta \nonumber\\
&= \Delta^{N} \gamma - \Delta^{N} y = \Delta^{N} \varepsilon_{\gamma}.
\end{align}

The remaining task is to recover $\varepsilon_{\gamma}$ from $\Delta^{N} \varepsilon_{\gamma}$. 
The reconstruction proceeds iteratively using the anti-difference operator $\mathrm{S}$ (Lines 5--6 of Algorithm~\ref{alg:noisy_RSoD}), recovering $\Delta^{N-1} \varepsilon_{\gamma}$ up to an unknown additive constant. This ambiguity arises because constant sequences are in the null space of the difference operator. The relationship is
\begin{equation}
(\Delta^{n-1} \varepsilon_{\gamma})[k] = (\mathrm{S} \Delta^{n} \varepsilon_{\gamma})[k] + \kappa_{n} \cdot 2\lambda,
\end{equation}
where $\kappa_{n} \in \mathbb{Z}$ must be determined at each step.

To resolve this integer ambiguity, we evaluate the difference (adapting Eqs.~(24)--(25) in~\cite{bhandari_unlimited_2021}):
\begin{align}
&(\mathrm{S}^2 \Delta^{n} \varepsilon_{\gamma})[1] - (\mathrm{S}^2 \Delta^{n} \varepsilon_{\gamma})[J+1] \nonumber\\
\quad \in &2\lambda \kappa_{n} J 
   + \left( 2 (T \Omega)^{n-2} \beta_{g} + 2^{n-1} \lambda \right) [-1,1] \\
\quad\subseteq  & 2\lambda J [\kappa_{n} - \varrho_{n}, \kappa_{n} + \varrho_{n}],
\end{align}
where
\begin{equation}
\varrho_{n} = \frac{1}{J} \left( \frac{\beta_{g}}{\lambda} + 2^{n-2} \right)
\end{equation}
This derivation mirrors~\cite{bhandari_unlimited_2021} but omits the constant $e$ due to Lemma~\ref{lem:FD-bound}. 

By setting $J = \lceil 4(\beta_g/\lambda + 2^{N-2}) \rceil$, we have $\rho_n \leq \frac{1}{4}$ for all $n \leq N$, ensuring $2\rho_n < \frac{1}{2}$. Therefore, the interval $[\kappa_{n} - \varrho_{n}, \kappa_{n} + \varrho_{n}]$ contains exactly one integer, allowing unique determination of $\kappa_n$ as
\begin{equation}\label{eq:calc_kappa}
\kappa_n = \left\lfloor \frac{(\mathrm{S}^2 \Delta^{n} \varepsilon_{\gamma})[1] - (\mathrm{S}^2 \Delta^{n} \varepsilon_{\gamma})[J+1]}{2J\lambda} + \frac{1}{2} \right\rfloor.
\end{equation}
Applying \eqref{eq:calc_kappa} for $n=N,N{-}1,\dots,2$ and antidifferencing with $2\lambda\mathbb{Z}$ alignment (Algorithm lines 5–10) yields $\gamma[k]+\eta[k]$ up to an additive $2m\lambda$.
\end{IEEEproof}

\section{Proof of Theorem~\ref{thm:noisy_N2_jitter}}\label{Asec:Jitter}
\begin{IEEEproof}
We focus on bounding $\Delta^2\gamma[k]$ for jittered sampling times 
$t_k=kT+\mu_k$, $|\mu_k|<\nu T$. Expanding around the uniform grid,
\begin{equation}
g(t_n) = g(nT) + \mu_n g^{(1)}(\xi_n),\quad n=k,\:k+1,\,k+2
\end{equation}
for some $\xi_n\in(nT,t_n)$ ($\mu_n>0$) or $\xi_n\in(t_n,nT)$ ($\mu_n<0$). Substituting the above into
\[
\Delta^2 \gamma[k] = g(t_k)+g(t_{k+2})-2g(t_{k+1}),
\]
we obtain
\begin{align}
\Delta^2 \gamma[k] &= \big(g(kT)+g((k+2)T)-2g((k+1)T)\big) \nonumber\\
&\quad + \mu_k g^{(1)}(\xi_k)+\mu_{k+2} g^{(1)}(\xi_{k+2})-2\mu_{k+1} g^{(1)}(\xi_{k+1})\nonumber
\end{align}
By the mean value theorem of forward difference,
\[
g(kT)+g((k+2)T)-2g((k+1)T) = T^2 g^{(2)}(\bar{\zeta}_k)
\]
for some $\bar{\zeta}_k\in(kT,(k+2)T)$. Therefore, applying Bernstein inequality produces
\[
|\Delta^2 \gamma[k]|
\;\le\; (T\Omega)^2 \|g\|_\infty 
+ (|\mu_k|+|\mu_{k+2}|+2|\mu_{k+1}|)\,\Omega\|g\|_\infty.
\]
Since each $|\mu_j|\le \nu T$ ($k\le j\le k+2$), this gives the improved bound
\[
|\Delta^2 \gamma[k]| \;\le\; (T\Omega)^2 \|g\|_\infty + 4\nu T \Omega \|g\|_\infty.
\]

For noisy measurements $y_\eta[k]=\mathcal{M}_\lambda(\gamma[k]+\eta[k])$,
we additionally have $|\Delta^2 \eta[k]|\le 4\|\eta\|_\infty$. 
Thus
\[
|\Delta^2 \gamma[k]+\Delta^2 \eta[k]|
\le \|g\|_\infty\big((T\Omega )^2+4\nu  T\Omega\big)+4\|\eta\|_\infty.
\]

If the RHS is strictly less than $\lambda$, the condition~
\eqref{eq:cond_N2_jitter_updated} holds and modulo cancellation applies:
\[
\mathcal{M}_\lambda(\Delta^2 \gamma[k]+\Delta^2 \eta[k])
=\Delta^2 \gamma[k]+\Delta^2 \eta[k].
\]
Note that as $N=2$, we 
\begin{align*}
&\:(\mathrm{S}^2 \Delta^{2} \varepsilon_{\gamma})[1] - (\mathrm{S}^2 \Delta^{2} \varepsilon_{\gamma})[J+1] \nonumber\\
=&\: \varepsilon_{\gamma}[1]-\varepsilon_{\gamma}[J+1]+2\lambda \kappa_{1}\\
\quad \in &\:2\lambda \kappa_{1} J 
   + \left( 2 \beta_{g} + 2 \lambda \right) [-1,1] %\\
%\quad\subseteq  &\: 2\lambda J [\nu_{2} - \varrho_{2}, \nu_{2} + \varrho_{2}],
\end{align*}
Thus, by choosing $J=4\beta_g/\lambda+2$, we can determine $\kappa_{1}$ uniquely. 
The remainder of the proof, unwrapping via anti-differencing and 
resolving the integer ambiguity, follows identically to the argument in 
Theorem~\ref{thm:noisy} or Theorem~3 in~\cite{bhandari_unlimited_2021}. Details are omitted.
\end{IEEEproof}

%\newpage
\bibliographystyle{IEEEtran}
\bibliography{refs}

\vfill

\end{document}